\documentclass{aa}  
\usepackage{graphicx,nicefrac}
\usepackage{txfonts}
\usepackage{hyperref}
\usepackage{color}
\usepackage{pdflscape}
\usepackage{multirow}
\usepackage{soul} 
\usepackage{algorithm}
\usepackage{algpseudocode}
\usepackage{float}

\DeclareRobustCommand{\teff}{T_{\mathrm{eff}}}
\DeclareRobustCommand{\logg}{\log g}
\DeclareRobustCommand{\mh}{\mathrm{[M/H]}}
\DeclareRobustCommand{\alph}{\mathrm{[\alpha/M]}}
\DeclareRobustCommand{\alphFe}{\mathrm{[\alpha/Fe]}}
\DeclareRobustCommand{\feh}{\mathrm{[Fe/H]}}
\DeclareRobustCommand{\kms}{\mathrm{km s}^{-1}}

\DeclareRobustCommand{\gaiadr}{\emph{Gaia} DR3}
\DeclareRobustCommand{\bprp}{$G_{BP}-G_{RP}$}
\DeclareRobustCommand{\logM0}{$\log (M_0/M_{\odot})$}
\mathchardef\mhyphen="2D
\renewcommand{\th}{$^{\mathrm{th}}$}

\begin{document}

   \title{New stellar age estimates using SPInS based on Gaia DR3 photometry and LAMOST DR8 abundances}
   \author{L. Casamiquela\inst{1}
          \and
          D. R. Reese\inst{2} \and
          Y. Lebreton\inst{2,3} \and
          M. Haywood\inst{1} \and
          P. Di Matteo\inst{1} \and
          F. Anders\inst{4,5,6} \and
          R. Jash\inst{2} \and
          D. Katz\inst{1} \and
          V. Cerqui\inst{1} \and
          T. Boin\inst{1} \and
          G. Kordopatis\inst{7}
          }

   \institute{
   GEPI, Observatoire de Paris, PSL Research University, CNRS, Sorbonne Paris Cité, 5 place Jules Janssen, 92190 Meudon, France \email{laia.casamiquela@obspm.fr}
   \and
    LESIA, Observatoire de Paris, Université PSL, CNRS, Sorbonne Université, Université Paris Cité, 5 place Jules Janssen, 92195 Meudon, France
    \and
    Univ Rennes, CNRS, IPR (Institut de Physique de Rennes) - UMR 6251, F-35000 Rennes, France
    \and
    Institut de Ci\`encies del Cosmos (ICCUB), Universitat de Barcelona (UB), Martí i Franquès, 1, 08028 Barcelona, Spain
    \and
    Departament de Física Quàntica i Astrofísica (FQA), Universitat de Barcelona (UB),  Martí i Franquès, 1, 08028 Barcelona, Spain
    \and
    Institut d'Estudis Espacials de Catalunya (IEEC), Esteve Terradas, 1, Edifici RDIT, Campus PMT-UPC, 08860 Castelldefels (Barcelona), Spain
    \and
    Université Côte d’Azur, Observatoire de la Côte d’Azur, CNRS, Laboratoire Lagrange, Nice, France
    }

   \date{Received ; accepted }

  \abstract 
   {Reliable stellar age estimates are fundamental for testing several problems in modern astrophysics, in particular since they set the time scales of Galactic dynamical and chemical evolution.}
   {In this study, we determine ages using only \gaiadr\ photometry and parallaxes, in combination with interstellar extinction maps, and spectroscopic metallicities and $\alpha$ abundances from the latest data release (DR8) of the LAMOST survey. In contrast with previous age estimates, we do not use spectroscopic effective temperatures or surface gravities, thus relying on the excellent precision and accuracy of the \emph{Gaia} photometry.}
   {We use a new version of the SPInS code \citep{Lebreton+2020} with improved features, including the on-the-fly computation of the autocorrelation time and the automatic convergence evaluation.}
   {We determine reliable age estimates for 35,096 and 243,768 sub-giant and main-sequence turn-off stars in the LAMOST DR8 low- and medium-resolution surveys with typical uncertainties smaller than 10\%. In addition, we successfully test our method on more than 4,000 stars of 14 well-studied open and globular star clusters covering a wide range of ages, confirming the reliability of our age and uncertainty estimates.} 
   {}


   \maketitle
%

\section{Introduction}

Determining ages for individual stars is among the most difficult problems in astrophysics because they cannot be measured directly. 
The most direct method to-date is nucleochronometry: radiodatation of the oldest meteorites provides a solid estimate of the solar system age and therefore of the age of the Sun \citep[e.g.][]{Chaussidon2007} while long half-life isotope abundances inferred from the spectra of a few very old metal-poor halo stars give a rather straightforward access to their age \citep[e.g.][]{Christlieb2016}.
For other stars, several age-dating methods are used in the literature to estimate ages \citep[see][for a review]{Soderblom2010}, even though no single method is valid for all ranges of ages or spectral types.

One can classify the different techniques into model-dependent ones, which make use of stellar evolutionary models, and empirical relations between the age and a given stellar observable.
In the latter type, the underlying physics of the empirical relations is usually not fully understood and needs to be calibrated on samples of stars with high-quality ages (obtained by model-dependent methods).
Some examples are gyrochronology which uses the empirical relation between a star's rotation period and its age \citep{Barnes2007}, or the correlation between activity or lithium abundance of F, G, K stars and age.
Another possibility, which has been recently boosted with the advent of large spectroscopic surveys, is to use the abundance-age relations which display linear dependencies with age partly explained by stellar and/or Galactic evolution.
This could be the case of the $\alphFe$-age relation \citep{Haywood+2013,Bensby+2014,Haywood+2015,Ciuca+2021,Katz+2021}, the CNO abundances (which are mostly driven by stellar evolution; \citealt{Masseron+2015,Lagarde+2017}), or the so-called chemical clocks (e.g. [Y/Mg] or [Sr/Mg] which are more driven by a smooth Galactic evolution of s-process elements; \citealt{Nissen2015, Nissen2016, TucciMaia+2016, Delgado-Mena+2019, Jofre+2020, Casamiquela+2021}).
As a generalisation, spectroscopic ages can be derived using supervised machine-learning techniques which take advantage of the full chemical information of the stars trained on high-precision datasets such as those based on asteroseismology \citep{Ciuca+2021, Hayden+2022, Anders+2023, Boulet2024}. 
This approach, however, has a significant dependence on which stellar population the training is performed on, and the age estimations are intrinsically entangled with Galactic chemical evolution, thus there is a potential redundancy when using them to study the chemical evolution of the Milky Way.

The methods which are considered most reliable are those which use stellar evolution models, i.e. asteroseismology or isochrone placing, for which the underlying physics is rather well understood \citep[but see][]{Lebreton2014a, Lebreton2014b} and which rely on the fewest assumptions.
Asteroseismology provides a precise way to constrain ages through the measurement of solar-like oscillations.
Until the advent of the PLATO mission \citep{Rauer2014} there are only limited samples of stars ($\lesssim 10,000$) with asteroseismic constraints and only in selected fields of the CoRoT \citep{ Baglin2006}, Kepler \citep{Borucki2010}, K2 \citep{Howell2014}, and TESS \citep{Ricker2015} missions \citep[see e.g. ][for a review]{Miglio2017}.
On the other hand, placing isochrones becomes \emph{easy} in open and globular clusters since they have many coeval mono-metallic members.
This allows a distribution fit over the whole range of mass is being fit, leading to a generally well-constrained situation\footnote{with the caveat of some globular clusters which do not have only mono-metallic members}.
That is why stellar clusters provide the primary benchmarks to study age-related properties.
In fitting a cluster age one has many stars distributed in mass in an HR diagram, and it is the full behaviour of the models over that mass range that is being fit, leading to a highly precise result overall.
For individual field stars the distribution fit is not possible, so isochrone placement becomes challenging because multiple isochrones can pass through a given point in the HR diagram given the degeneracies in the different observables and the underlying photometric uncertainties.
Indeed, this method can only be applied with good precision ($\lesssim 15\%$) for individual stars near the main sequence turn-off or in the subgiant branch, where evolutionary models of different masses separate in the HR diagram, thus providing less degeneracy \citep{Jorgensen2005}.
Therefore, both methods, asteroseismology and isochrone placement, are complementary in terms of spectral types of stars for which precise ages can be obtained.

The advent of \emph{Gaia} with its precise photometry and parallaxes for a vast number of stars, has allowed massive stellar parametrisation.
With \gaiadr\ \citep{GaiaCollaboration+2023}, stellar parameters for 471 million sources were estimated from low-resolution BP/RP spectra \citep{Andrae+2023}, together with the chemo-physical parametrisation for 5.6 million stars estimated from the Radial Velocity Spectrometer (RVS) data \citep{RecioBlanco+2023}.
The module FLAME then produced luminosities, radii, masses and ages among other parameters, for 284 million stars in \gaiadr. Ages, computed mainly using atmospheric parameters ($\teff$, $\logg$, and $\mh$) and absolute magnitude as inputs, were produced for half of the sample \citep{Creevey2023}.

Many other studies in the literature have used \emph{Gaia} in combination with other external data to produce more precise ages (e.g. \citealt{McMillan+2018, Sanders+2018, Wang2023, Stone-Martinez2024}).
For instance, recently, \citet{Xiang+2022} computed ages for subgiant and main sequence turnoff stars which were observed as part of the seventh data release of the large-scale spectroscopic survey LAMOST \citep{Cui2012, Zhao2012}.
They made use of absolute magnitudes in the $K$ band, and spectroscopic estimates of $\teff$, $\feh$ and $\alphFe$ abundances from the DD-payne pipeline in LAMOST DR7.
\citet{Queiroz+2023} recently used the {\tt StarHorse} code \citep{Queiroz+2018, Anders+2022} to derive stellar ages for main-sequence turnoff and subgiant branch stars for millions of stars observed by several spectroscopic surveys, including e.g. APOGEE \citep{Majewski+2017}, GALAH \citep{DeSilva+2015}, and {\it Gaia} RVS. 
In this case, additionally to \gaiadr~data and spectroscopically determined atmospheric parameters (effective temperature, surface gravity, metallicity), it uses infrared photometry from 2MASS and AllWISE to infer stellar astrophysical parameters, including ages.
\citet{Kordopatis+2023} determined Bayesian isochrone age estimates for stars observed by {\it Gaia}'s Radial Velocity Spectrograph (RVS).
They used GSP-Spec \citep{RecioBlanco+2023} calibrated atmospheric parameters, 2MASS and {\it Gaia}-EDR3 photometry, and parallax-based distances to compute the ages, initial stellar masses, and reddenings for 5 million stars with spectroscopic parameters in {\it Gaia}-DR3.

This study aims to determine ages using only \gaiadr\ photometry ($G$,$G_{BP}$,$G_{RP}$) and parallax, in combination with recent interstellar extinction 3D maps, and chemical abundances from the latest data release (DR8) of  LAMOST.
In contrast with previous age estimates, here we do no make use of a particular stellar parametrisation (e.g. spectroscopic $\teff$ or $\logg$) because the information of the HR diagram is provided through the absolute \emph{Gaia} magnitudes only.
Indeed, one of the strong points of the \emph{Gaia} mission is to deliver very precise and accurate photometry and parallaxes which makes it possible to date stars in a HR diagram with the lowest possible uncertainties.
In turn, spectroscopic information on the metallicity and chemical composition ($\alpha$ abundance in particular) allows us to break some degeneracies of isochrone fitting.
We use the recent code SPInS \citep{Lebreton+2020}, which uses a Bayesian framework with a MCMC sampler coupled with an interpolation scheme, to parameterise stars using stellar evolutionary models.

The paper is structured as follows: Section \ref{sec:method} explains the basics of our method (for a detailed paper we refer the reader to \citealt{Lebreton+2020}).
Section \ref{sec:data} describes the the selection of LAMOST and \gaiadr~data used in this paper.
In Sect. \ref{sec:val} we present a detailed comparison with open and globular clusters to validate our method, and in Sect. \ref{sec:results} and \ref{sec:discussion} we show and discuss the results of applying our method to the field star sample described in Sect. \ref{sec:data}.
Finally, we conclude the paper in Sect. \ref{sec:conclusions}.

\section{The method}\label{sec:method}
We aim to obtain stellar parameters, in particular ages, of a large sample of stars (selected as described in Sect.~\ref{sec:data}).
To this end, we use the code Stellar Parameters INferred Systematically \citep[SPInS][]{Lebreton+2020}, a public {\tt python} pipeline\footnote{\url{https://gitlab.obspm.fr/dreese/spins}} which takes different types of inputs (e.g. photometric, spectroscopic, interferometric and/or averaged asteroseismic) to provide age, mass, and radius (among others) of a star, relying on a grid of evolutionary tracks. 
In brief, the code works in a Bayesian framework to provide the posterior probability distribution function (PDF) of the inferred stellar parameters from a set of observational constraints, a grid of stellar models (see Sect. \ref{sec:basti}), and a set of priors.
The PDF is sampled using an MCMC solver based on the \emph{emcee} python package \citep{Foreman+2013}, coupled with a versatile interpolation scheme for the stellar models (see Sect. \ref{sec:mcmc}).

\subsection{Stellar grids and priors}\label{sec:basti}
We use the BaSTI\footnote{\url{http://basti-iac.oa-abruzzo.inaf.it}} grid of stellar models  \citep{Hidalgo2018}, calculated for a solar-scaled heavy elements distribution and updated input physics, including atomic diffusion of helium and metals, overshooting of convective cores, and mass loss.
The corresponding solar mixture is that of \citet{Caffau2011} complemented by \citet{Lodders2010}.
Evolutionary tracks are provided for  a set of masses and metallicities (see below), with a helium abundance derived assuming an helium to metal enrichment ratio $\Delta Y/\Delta Z=1.31$ \citep[see][for more details]{Hidalgo2018}.
For each stellar model of given age, mass, and chemical composition, the luminosity and effective temperature are provided, as well as Gaia EDR3 magnitudes $M_G, M_{G_\mathrm{BP}}, M_{G_\mathrm{RP}}$.
On SPInS website, the previous BaSTI stellar evolution tracks \citep{Pietrinferni+2004} are provided in a format directly readable by SPInS. 
However, any stellar evolution grid, calculated with any stellar evolution code, can be used if it is written in SPInS input format.

For the sake of homogeneity, in order to use always the same models for all stars, we do not use the $\alpha$-enhanced model grid, but we use the solar-scale grid scaling the input metallicities to mimic the $\alpha$-element enrichment.
This is done via the commonly used relation derived by \citet{Salaris+1993}\footnote{In \citet{Salaris+1993} the solar-scaled $Z$ is re-scaled as $Z_{\alpha}$ according to $Z_{\alpha} = Z \times (0.638 \times 10^{\alph} + 0.362)$}.

The whole BaSTI grid contains a total of 1,120 evolutionary tracks of initial mass in the range  $M_0\in[0.1M_{\odot},15M_{\odot}]$  and metallicity [M/H] $\in[-3.197, +0.3]$.
The grid covers all evolutionary stages from pre main sequence up to either first thermal pulses on the AGB or C-ignition, or to the age of the universe, depending on the mass.
In the grid prepared for SPInS, we have excluded high masses $M>10 M_\odot$, the pre-main sequence parts of each track ($t_{adim}<0.05$) as well as the phases beyond the RGB tip, given the low probability of having pre-main sequence, very evolved or massive stars in our sample selected as described in Sect.~\ref{sec:data}.
$t_{adim}$ is an adimensional age parameter, which goes from 0 to 1, whose value depends on the evolutionary stage, and is thus homogeneous for all masses.  
The used stellar evolutionary model grid has a total of 978,732 points.

Interpolation on the grid is done in the (\logM0, $\mh$, $t_{adim}$) space, at each MCMC iteration. On the one hand, this involves the interpolation between evolutionary tracks, which is coded as a linear barycentric interpolation on a simplex defined by a Delaunay tesselation on the grid of models.
On the other hand, the interpolation along the evolutionary track is linear between adjacent points, performed on $t_{adim}$, with the purpose to combine models at the same evolutionary stage.
Then, a transformation to physical age ($t$) is implemented in SPInS for each MCMC step to provide a sample on physical age.
We refer the reader to \citet{Lebreton+2020} for more technical details about the interpolation scheme.

SPInS allows the use of priors on the grid parameters  such as the initial mass function, the metallicity distribution function or the star-formation rate.
In the case of this work, we do not impose any strict prior on metallicity, mass or age to avoid biases in the statistical interpretation of the resulting trends.
This is motivated by the fact that overall systematic offsets can be present in the derived ages depending for instance on the set of evolutionary tracks or the photometric transformations.
This means that ages larger than the age of the Universe are allowed, as much as this is allowed by the evolutionary tracks.

\subsection{Configuration of the MCMC sampler and evaluating the convergence}\label{sec:mcmc}

There are several options concerning the MCMC sampler integrated in SPInS that need to be fixed in the code on a case-by-case basis.

Firstly, we choose to do the initialisation of walkers based on a Gaussian distribution centred around the best-fitting model.
This is a convenient option to reach quick convergence, since it already places the starting points near a preferred position, although later the walkers can move and explore more distant points.

Other free parameters such as the number of walkers, burn-in and production steps have a large dependence on the complexity of the posterior and need to be properly set to allow a correct exploration of the parameter space.
This is a crucial issue because, if the PDF is multimodal (as is often the case in an HR diagram), it can be tough to sample it with a standard MCMC, and can result in biased solutions in some regions of the HR diagram.
A possible solution to this problem is to use a parallel-tempering ensemble MCMC which runs in a modified posterior given by a transformation of the likelihood as:
\begin{equation}
    \mathcal{L}'(x) = \mathcal{L}(x)^{(1/T)},
\end{equation}
where $T$ is a parameter usually called "temperature" \citep{Vousden+2015}.
Using higher $T$ values allows MCMC chains to explore the parameter space more easily because the likelihood is flatter and broader.
The parallel-tempered MCMC is implemented to be used in SPInS via the package \emph{ptemcee}\footnote{\url{https://github.com/willvousden/ptemcee}}.
In our case, we have seen that this strategy dramatically improves the convergence of the chains.
However, the usage of parallel tempering implies that, for a fixed number of iterations, the computation times increase with the number of temperatures, reaching several minutes per star.
Therefore we found a need to evaluate the convergence on-the-fly for each star, in order to accelerate the computational run.

Evaluating the convergence of the walker chains is an essential step in an MCMC analysis.
Even though it is formally impossible to guarantee the convergence of an MCMC sampler, there are some diagnostic tools to evaluate if we are obtaining an accurate approximation of the PDF.
We choose to use a criterion based on the integrated autocorrelation time ($\tau$), whose basic idea is that the chain has sampled long enough when the walker has traversed the high-probability parts of the parameter space many times in the length of the chain.
Following the discussion of \citet{Hogg+2018}: a small value of $\tau$ compared with the length of the chain can be used as a sign of convergence.

In this work, we have implemented in SPInS a robust on-the-fly computation of the autocorrelation time\footnote{There are different ways of computing autocorrelation times, we follow the prescriptions described in \url{https://dfm.io/posts/autocorr/}} done while sampling the MCMC every $n_{\mathrm{steps}}$.
We consider that the sample has reached convergence when the mean of $\tau$ in the three main dimensions of the grid (\logM0, $\mh$, $t$) reaches: $\tau < N/100$, where $N$ is the number of steps that the chain has sampled.
With the achievement of the convergence criteria, we stop the MCMC sampler and draw the statistical analysis of the resulting PDF.
This strategy is particularly useful for running SPInS in a massive way because it allows us to save computation time once convergence is reached, and provides an automatic evaluation of the goodness of the PDF when the MCMC has finished.

A new version of SPInS with improved features including the on-the-fly computation of the autocorrelation time and the automatic convergence evaluation used in this paper, is publicly available.
Our tests show that with 10 temperatures, 10 walkers and 1,000 burn-in steps, most of the stars in our sample (main sequence turnoff and subgiants, see Sect.~\ref{sec:data}) reach convergence in around 2,000 production steps.
With this configuration on an Apple M1 processor, SPInS takes around 20 seconds per star using the native SPInS parallelisation on 2 processes.
On top of this, in this study we parallelize among 20 nodes the sample of main sequence turnoff and subgiant stars detailed in Sect.~\ref{sec:MSTOSGB}.

\section{Data selection}\label{sec:data}

\begin{figure*}[htp]
\centerline{\includegraphics[width=0.8\textwidth]{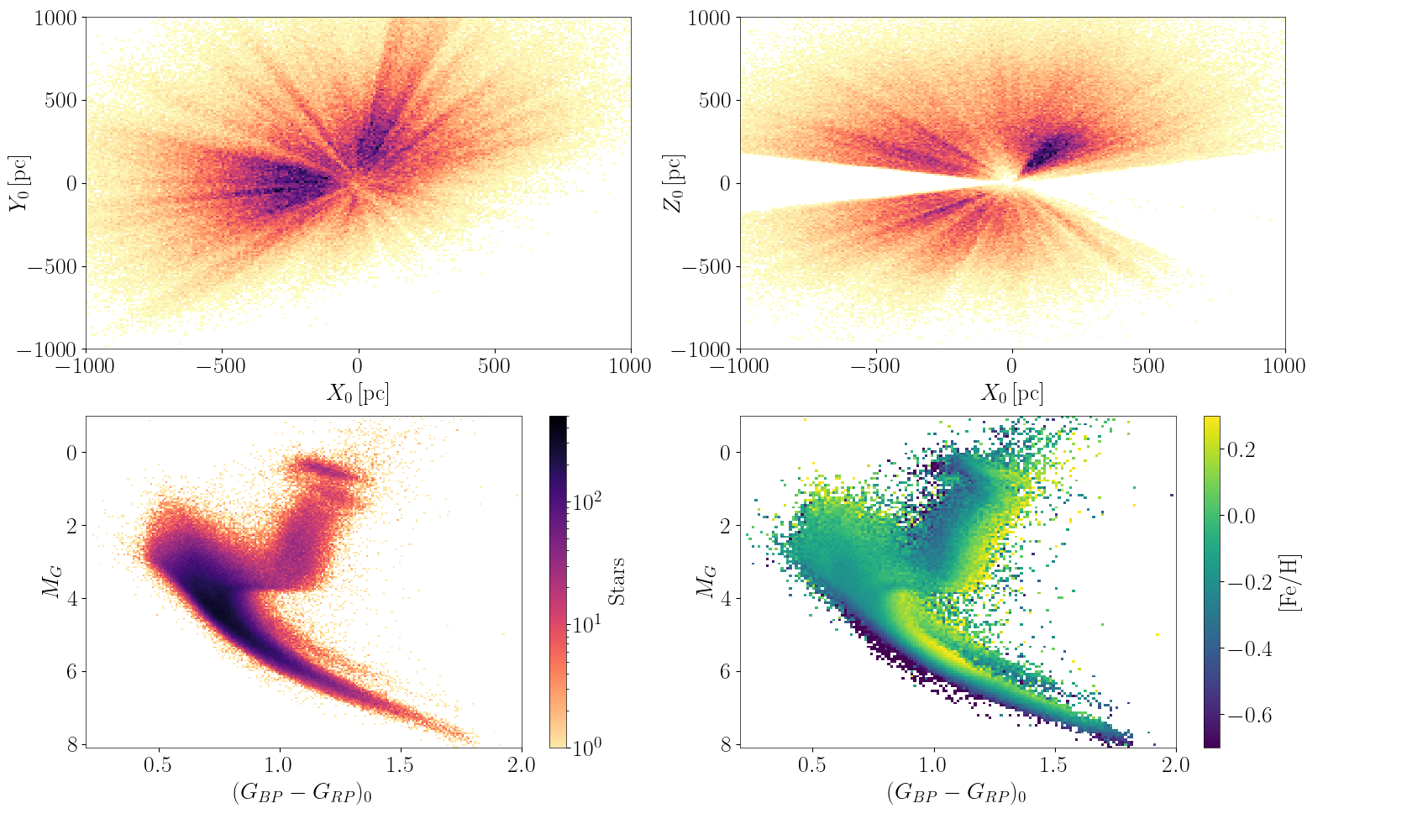}}
\caption{Distribution of the initial sample of 352k stars in the LAMOST DR8 LRS sample obtained from the procedure indicated in Sec.~\ref{sec:data}. Top: Galactic X-Y and X-Z distributions coloured by stellar density. The Sun is at (0,0). Bottom: Intrinsic CMD coloured by density (left), and binned with colours representing mean $\feh$\ (right).}
\label{fig:init_sample}
\end{figure*}

SPInS is very flexible in terms of input observational constraints; any parameter included in the grid of evolutionary models can essentially be used as a constraint.
In this paper, with the exception of spectroscopic metallicity estimates, we exclusively use photometric and astrometric constraints: $M_G$ and (\bprp)$_0$, obtained from \gaiadr\ photometry and parallax, coupled with a reddening estimate from the recent three-dimensional extinction maps provided by \citet{Vergely+2022}.
These maps are based on the inversion of large spectroscopic and photometric catalogues including \gaiadr, 2MASS, and AllWISE.

To build our final sample we perform an initial query in the \gaiadr\ archive selecting stars brighter than magnitude $G=18$ and in a 3 kpc region around the Sun.
Being able to retrieve accurate absolute magnitudes and intrinsic colours is essential for the isochrone placement method.
We perform cuts on the relative uncertainties in the parallax and in the three magnitudes.
We perform cuts on Galactic latitude to avoid high extinction regions in the Galactic plane, where the extinction maps are less accurate.
Additional cuts are applied on the normalised unit weighted error (ruwe), the percentage of successful Image Parameter Determination (IPD)-windows with more than one peak (ipd\_frac\_multi\_peak), and the amplitude of the IPD goodness–of–fit (GoF), to avoid binary/multiple stars.
The complete query is written below:

\begin{verbatim}
SELECT * FROM gaiaedr3.gaia_source WHERE 
parallax_over_error > 10 AND
phot_g_mean_flux_over_error>50  AND 
phot_rp_mean_flux_over_error>20 AND 
phot_bp_mean_flux_over_error>20 AND 
phot_g_mean_mag < 18 AND 
parallax > 0.33 AND
ipd_frac_multi_peak < 2 AND
ipd_gof_harmonic_amplitude < 0.1 AND
ruwe<1.1 AND (b<-10 OR b>10) AND dec>-10 
\end{verbatim}

We have additionally performed a cut in declination because we later crossmatch the sample with LAMOST (see Sect.~\ref{sec:LAMOST}), which is entirely contained above -10$^{\circ}$.
This query gives a total of 23 million stars.

\subsection{Building the intrinsic colour-magnitude diagram}\label{sec:data-CMD}

For the computation of interstellar absorption, we choose the map from \citet{Vergely+2022} which covers a volume of 3 kpc x 3 kpc x 800 pc around the Sun at a resolution of 10 pc.
The individual $A_V$ values per star are computed with a linear 3D interpolation of the absorption density towards the line of sight, followed by an integration along the line of sight.
Some of the stars are located outside the map at $|Z|>800$ pc. Accordingly, we consider the absorption to be zero at higher $|Z|$ values.
Absolute magnitudes are computed using an inversion of the parallax. 

We applied zero-point corrections on Gaia DR3 parallaxes as recommended by \citet{Lindegren21}.
In all cases, we obtained the apparent magnitudes $G, BP, RP$ from the photometric fluxes and  corrected for \emph{Gaia} colour zero points following  \citet{Riello2021}.
To correct for interstellar extinction, we used the empirical relations provided by \citet{Danielski2018} with coefficients updated for \emph{Gaia} DR3\footnote{\url{https://www.cosmos.esa.int/web/gaia/edr3-extinction-law}}.
For instance, the extinction coefficient for the $G$ band is $k_G((G_{BP}-G_{RP})_0, A_0) = A_G/A_0$ where  $(G_{BP}-G_{RP})_0$ is the star's intrinsic colour, and $A_G$ and $A_0$ are the extinction in the $G$-band and at $\lambda_0=550$ nm, respectively.
For stars having $(G_{BP}-G_{RP})\leq -0.06$ which is the limit of validity of \citet{Danielski2018}'s formulae, we used \citet{Wang+2019}'s law.
For $A_0$, we used the value of $A_V$.
The used extinction maps do not have detailed uncertainty values, thus, we warn the reader that the derived uncertainties on the absolute magnitudes are a lower limit.

\subsection{Metallicities and $\alpha$ abundances}\label{sec:LAMOST}

The previous sample is finally crossmatched with the DR8 catalogue for A, F, G, K stars of the LAMOST DR8 spectroscopic survey to use the $\feh$\ and $\alphFe$ abundances as additional constraints, which significantly breaks degeneracies to get the physical parameters from evolutionary tracks.

LAMOST DR8 contains:

\begin{itemize}
    \item a spectroscopic parameter catalogue of 6.6 million stars from the low-resolution survey (LRS) which provides atmospheric parameters, radial velocities, iron and $\alpha$-element abundances obtained with a spectral resolution of 1,800 in the wavelength range of 3700-9000 \AA.
    \item a spectroscopic parameter catalogue of 1.2 million stars coming from the medium resolution (7,500) survey (MRS) which provides, additionally, overall $\alpha$-element abundances and individual abundances for certain elements. In this case, two sets of parameters and abundances are listed, one coming from the LASP pipeline \citep{Wu+2014}, and another set coming from a convolutional neural network (CNN), which up to DR8 was called Data-Driven Payne (DD-Payne). \citet{Soubiran+2022} did a thorough comparison of the $\feh$\ results among spectroscopic surveys, finding that DD-Payne for LAMOST DR5 tends to give larger biases than LASP in the metal-poor regime ($\feh<-1$), when compared to higher resolution studies. In this work we have used both estimations of $\feh$\ and $\alphFe$ (LASP and CNN) to compute two sets of MRS ages.
\end{itemize} 

Cuts on the $\feh$\ uncertainty of $<0.05$   
and the radial velocity uncertainty of $<5\,\kms$ were applied to both samples to filter low-quality values and problematic spectra.
Once joined with the \emph{Gaia} selection (see previous subsection), we end up with a total of 490,233 stars in the LRS sample and 71,473 in the MRS sample.

Figure~\ref{fig:init_sample} shows the heliocentric $X-Y$ and $X-Z$ distributions of the LAMOST DR8 LRS sample, together with the corresponding colour-magnitude diagram (CMD) corrected by extinction.
The intrinsic CMD in the bottom-left panel, coloured by density, exhibits a thin main sequence, which also makes visible an equal-mass binary sequence in the faint magnitudes.
Clear turnoff, subgiant and red giant branches can be seen, together with a prominent red clump of helium-burning stars.
Below the red clump, there is a sign of the red giant branch bump, caused by a discontinuity in the luminosity of hydrogen shell-burning stars \citep[see e.g.][]{King1985}.
The bottom-right plot of Fig. \ref{fig:init_sample} shows the same intrinsic CMD, but now binned and coloured depending on mean $\feh$, which highlights a gradient with metallicity in the horizontal direction, where metal-rich stars tend to be redder, and metal-poor ones bluer.
This is particularly remarkable for the main sequence, red giant branch and red clump, as expected from stellar evolution models.
The secondary red clump is visible as a small compact group towards the bluer and fainter part of the main red clump.
This corresponds to more massive red clump stars, which are probably younger \citep{Girardi1999}, and this is consistent with the plot since one would expect them to be overall more metal-rich.

\begin{figure*}[htp]
\centerline{\includegraphics[width=0.99\textwidth]{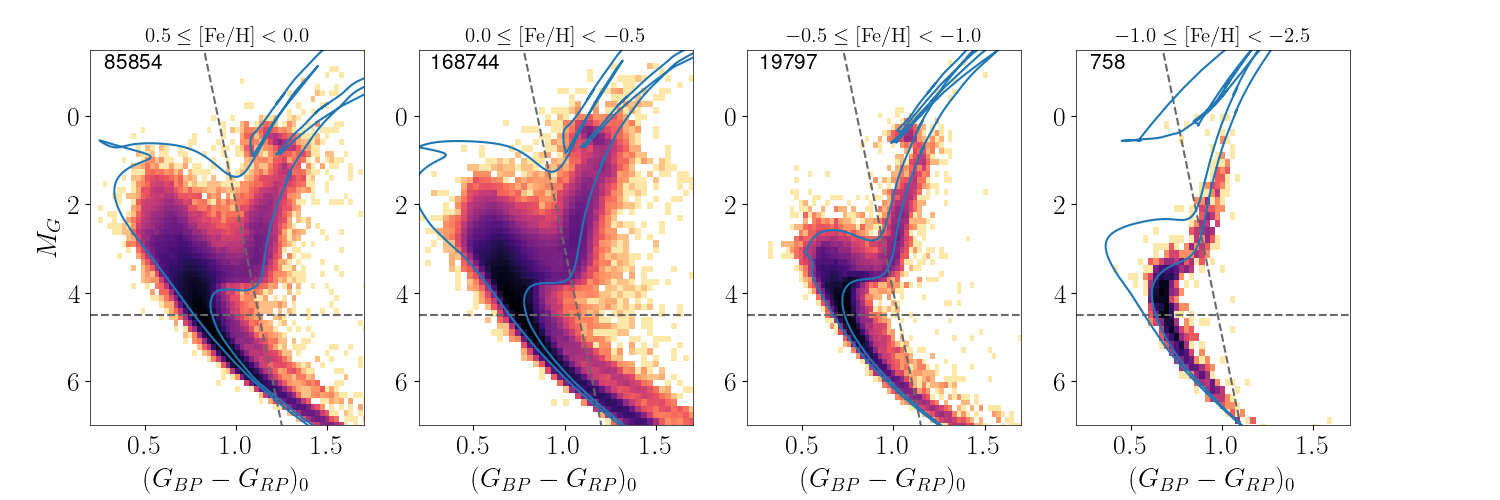}}
\centerline{\includegraphics[width=0.99\textwidth]{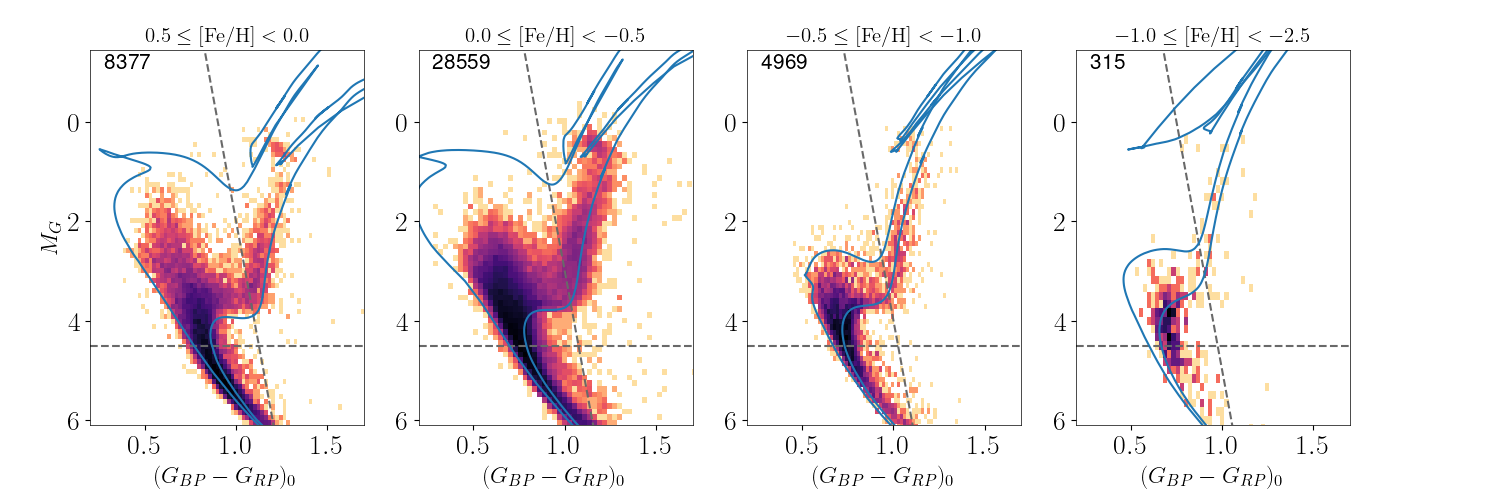}}
\caption{Intrinsic CMD of the initial sample of LRS (top) and MRS (bottom) stars, divided in four $\feh$\ bins, coloured by the density of stars (logarithmic scale).
The number of stars is plotted in each panel.
The grey lines represent the cuts done to select MSTO and SGB, as explained in Sec.~\ref{sec:MSTOSGB}.
To guide the eye, we additionally plot in blue two isochrones representative of each $\feh$\ bin at two different ages.}
\label{fig:MSTOSGB}
\end{figure*}

\begin{figure}[htp]
\centerline{\includegraphics[width=0.5\textwidth]{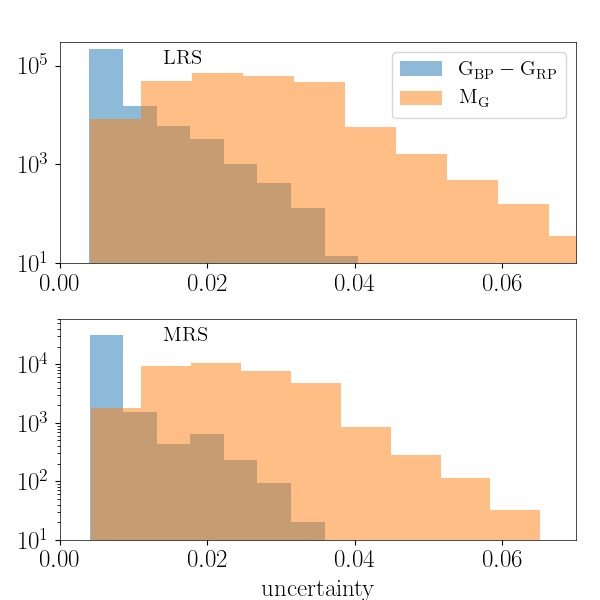}}
\caption{Distribution of uncertainties in absolute magnitude and colour for the selection of MSTO and SG stars in the MRS (bottom) and LRS (top) samples.}
\label{fig:distrib_unc}
\end{figure}

\subsection{Selection of main sequence turnoff and subgiant stars}\label{sec:MSTOSGB}

It is widely known (e.g. \citealt{Pont2004, Takeda2007}) that ages for individual stars derived by isochrone fitting are only reliable for the main sequence turn-off (MSTO) and subgiant branch (SGB) regimes because these positions in the Hertzsprung-Russell diagram have fewer degeneracies in the evolutionary models for different masses.
For stars in the red-giant branch and the main sequence, SPInS typically obtains very flat or ill-defined PDFs which are uninformative of the age of the star \citep[see tests done with synthetic stars in ][]{Lebreton+2020}.

Thus, we select the MSTO and SGB regions in bins of $\feh$\ as shown in Figure~\ref{fig:MSTOSGB}, similar to the MSTO/SGB selection done in \citet{Queiroz+2023}.
The faint end of the MSTO is selected with a horizontal cut at magnitude $M_G=4.5$, and the redder limit of the SGB is selected with a linear function $M_G = 20\cdot ($\bprp$)_0 + b_i$, where $b_i = [18,17,16,15]$ for the metallicity bins $\feh=(0.5,0.0),(0.0,-0.5),(-0.5,-1.0),(-1.0,-2.5)$, respectively.

The selection gives 243,768 stars in the LRS and 35,096 stars in the MRS.
Both samples consist of stars with an exquisite photometric and astrometric quality with parallax errors of 1\% in mean.
This provides very good uncertainties in the absolute photometry which are are in general smaller than 0.04 in $M_G$, and of the order of 0.01 in $G_{BP}-G_{RP}$, see Fig.~\ref{fig:distrib_unc}.
Metallicity errors have values of 0.03 dex in mean for both samples.

\section{Validation sample: open and globular clusters}\label{sec:val}

We validate the method of determining ages of individual field stars using open and globular clusters.
Even though they are also model dependent, star cluster ages are among the best anchors for validating age estimates of field stars, since they represent, in general, mono-age, mono-metallicity populations and thus it is possible to do a distributed isochrone fit along all mass range.
Indeed, the observed colour-magnitude diagrams of star clusters serve as important calibrators for stellar evolutionary models.

For our case, we select a set of validation stars in stellar clusters over a wide range of ages (see Table~\ref{tab:clusterdata}), similarly selected as our sample of field stars.
We then compute ages of the individual cluster member stars, independently of the cluster they belong to, to then compare them with cluster ages obtained in the literature.
Most of the literature cluster ages are obtained with a distributed isochrone fit on \emph{Gaia} photometry, using different stellar isochrones and fitting methods.
We also include ages determined using eclipsing binaries, asteroseismic ages for giants, and from the white dwarf cooling sequence, as described in the following subsection.

\begin{table*}[]
    \centering
    \caption{Table of clusters with their properties taken from the literature. Distances to OCs correspond  to the values derived by \citet{Cantat-Gaudin+2020}, except for the globular cluster NGC 6397 for which we use distance and extinction derived by \citet{Baumgardt+2021}. References for the extinctions ($A_V$) and ages are indicated with superscript numbers, and for metallicites are indicated with superscript letters.
    We include the ages determined in this work (see Sect.~\ref{sec:clusterres}, using the median and 16/84 quantiles, and the mode of the distributions), and the number of stars used ($N_{\mathrm{stars}}$).}
    \begin{tabular}{lccrrrrr}
    \hline

        Cluster & Distance & $A_V$    & Age   & $\feh$ & Age$_\mathrm{SPInS,mode}$ & Age$_\mathrm{SPInS}$ & $N_{\mathrm{stars}}$\\
            & (pc)     & (mag) & (Myr) & (dex) & (Myr) & (Myr) & \\
    \hline
Melotte 22 &   128 &  0.18 $^1$ &   77 $^1$,109 $^2$,125 $^4$               &  $ 0.051 \pm 0.078$ $^{ii}$  &   150 & $230  ^{+ 161}_{-  68}$ &   6 \\
  NGC 1039 &   534 &  0.24 $^2$ &   131 $^1$,250 $^2$                       &  $-0.046 \pm 0.092$ $^{ii}$  &   350 & $386  ^{+ 822}_{- 111}$ &  20 \\
  NGC 3114 &  1021 &  0.27 $^1$ &   144 $^1$,130 $^3$,158 $^4$              &  $ 0.050 \pm 0.060$ $^i$     &   350 & $338  ^{+ 193}_{- 108}$ & 133 \\
  NGC 2287 &   688 &  0.06 $^3$ &   169 $^1$,250 $^3$,200 $^4$              &  $-0.110 \pm 0.010$ $^i$     &   650 & $1071 ^{+1570}_{- 573}$ & 170 \\
  NGC 6475 &   285 &  0.30 $^1$ &   223 $^1$,346 $^2$,251 $^4$              &  $-0.012 \pm 0.050$ $^{iii}$ &   350 & $592  ^{+ 840}_{- 243}$ &  87 \\
  NGC 2516 &   423 &  0.22 $^2$ &   239 $^1$,302 $^2$,178 $^4$              &  $ 0.050 \pm 0.110$ $^i$     &   450 & $457  ^{+ 129}_{- 113}$ &  21 \\
  NGC 3532 &   498 &  0.07 $^2$ &   350 $^1$,398 $^2$,350 $^3$,354 $^4$     &  $-0.034 \pm 0.012$ $^{iii}$ &   650 & $1082 ^{+1432}_{- 485}$ & 408 \\
  NGC 2447 &  1018 &  0.11 $^3$ &   549 $^1$,549 $^2$,560 $^3$,562 $^4$     &  $-0.050 \pm 0.010$ $^i$     &   650 & $774  ^{+ 405}_{- 158}$ & 118 \\
  NGC 2632 &   183 &  0.08 $^2$ &   676 $^1$,708 $^2$,630 $^4$              &  $ 0.118 \pm 0.014$ $^{iii}$ &   750 & $829  ^{+ 288}_{- 158}$ &  45 \\
  NGC 5822 &   854 &  0.39 $^1$ &   912 $^1$,890 $^3$,891 $^4$              &  $ 0.080 \pm 0.080$ $^i$     &   950 & $1074 ^{+ 273}_{- 178}$ & 113 \\
  NGC 6819 &  2765 &  0.40 $^1$ &   2238 $^1$,1995 $^4$,2400 $^5$,2200 $^6$ &  $-0.050 \pm 0.010$ $^{iii}$ &  2950 & $3151 ^{+ 864}_{- 627}$ & 136 \\
  NGC 2682 &   889 &  0.12 $^3$ &   4265 $^1$,3467 $^2$,3630 $^3$,3630 $^4$ &  $-0.075 \pm 0.007$ $^{iii}$ &  4450 & $5166 ^{+1297}_{-1022}$ & 124 \\
  NGC 188  &  1698 &  0.26 $^2$ &   7079 $^1$,5495 $^2$,7585 $^4$,6000 $^7$ &  $-0.030 \pm 0.015$ $^{iii}$ &  6750 & $7412 ^{+1312}_{-1091}$ & 143 \\
  NGC 6397 &  2488 &  0.56      &  12600 $^8$,13000 $^9$,12800 $^{10}$      &  $-1.990 \pm 0.010$ $^{iv}$  & 13450 & $13292^{+ 764}_{- 943}$ &  83 \\
\hline
    \end{tabular}
    {\flushleft $^1$\citet{Cantat-Gaudin+2020}, $^2$\citet{GaiaCollaborationB+2018}, $^3$\citet{Tsantaki+2023}, $^4$\citet{Netopil+2022}, $^5$\citet{Brewer+2016}, $^6$\citet{Rodrigues+2017}, $^7$\citet{Meibom+2009}, $^8$\citet{Correnti+2018}, $^9$\citet{VandenBerg+2013}, $^{10}$\citet{Torres+2015}.
    $^i$\citet{Netopil+2016}, $^{ii}$\citet{Zhong+2020}, $^{iii}$\citet{Casamiquela+2021}, $^{iv}$\citet{Carretta+2009}
    
}
    \label{tab:clusterdata}
\end{table*}

\subsection{Cluster and star selections}\label{sec:clustersel}
We first select a suitable sample of clusters, starting from the open cluster catalogue of \citet{Cantat-Gaudin+2020} which we crossmatch with \gaiadr.
We restrict the sample of stars using the filters on photometric and astrometric quality from the \emph{Gaia} catalogue specified in Sec.~\ref{sec:data}, as done for the field stars.
To allow a better statistics in the comparison, we only keep the clusters which have a significant number of selected stars ($>500$), and which have an age and extinction determination in the \citet{Cantat-Gaudin+2020} catalogue.
This results in a sample of 9k stars in 13 open clusters.
We also search for $\feh$\ determinations of the selected clusters in the literature \citep{Casamiquela+2021, Netopil+2016}, prioritising those coming from LAMOST \citep{Zhong+2020} for consistency with Sect.~\ref{sec:data}.

We noticed that for some open clusters (NGC 1039, NGC 2287, NGC2516, NGC 3532, NGC2447, NGC 2632, NGC 2682, NGC 188), the extinction values derived by the automated procedure from \citet{Cantat-Gaudin+2020} are underestimated by 0.05 up to 0.1 mag with respect to other literature values.
In particular, \citet{GaiaCollaborationB+2018} and the recent determinations from \citet{Tsantaki+2023} provide very coherent values among them for the clusters in common, and also compared to previous literature studies.
The assumed extinction has an impact in the derived ages, so we decide to use the $A_V$ values from \citet{Tsantaki+2023} and \citet{GaiaCollaborationB+2018} for the eight mentioned clusters.

We list in Table~\ref{tab:clusterdata} the clusters' physical characteristics from the literature used in this work, including mean distances, absorption ($A_V$) and $\feh$ abundances.
We also include a non-exhaustive list of literature ages for these clusters, taken mainly from the studies of \citet{Cantat-Gaudin+2020}, \citet{GaiaCollaborationB+2018} and \citet{Tsantaki+2023}.
We find it relevant to include in this list ages coming from eclipsing binaries for NGC 6819 \citep{Brewer+2016} and NGC 188 \citep{Meibom+2009}, as well as asteroseismic ages for giants obtained by \citet{Rodrigues+2017} for NGC 6819.

As a test for old ages we use globular cluster members from the catalogue from \gaiadr\ \citet{Vasiliev+2021}.
We apply a strategy analogous to the case of open clusters, filtering stars according to their photometric and astrometric quality.
We additionally require the cluster to be closer than 3 kpc and to have a well populated turnoff with high probability members ($>0.7$) at magnitude $G<18$.
This selection yields only two clusters, NGC 6397 and NGC 6121 (M 4).
We discard NGC 6121 for this validation, since it is highly extincted due to its location behind the Upper Scorpius star forming region.
Its CMD shows a very broad main sequence and main sequence turnoff, pointing to a significant differential reddening.
Therefore, we keep as the only suitable globular cluster NGC 6397, which is located at $2.488\pm0.019$ kpc with an overall reddening of $E(B-V) = 0.18$ \citep{Baumgardt+2021}.
We adopt an $\feh=-1.99\pm0.01$ and an $\alphFe=0.46\pm0.04$ (from [Mg/Fe]) based on the mean values from a high spectroscopic resolution analysis of 13 member stars \citep{Carretta+2009}.
We also list in Table~\ref{tab:clusterdata} its properties, including several age determinations from the literature obtained with isochrone fitting \citep{Correnti+2018,VandenBerg+2013}, as well as an age derived from its white dwarf cooling sequence \citep{Torres+2015}.

We obtain an intrinsic colour-magnitude diagram for the cluster sample in the same way as for field stars, described in Sect.~\ref{sec:data-CMD}.
We applied the zero-point corrections on \gaiadr\ parallaxes to obtain the distances for the individual stars of the open clusters.
For the globular cluster, we did not apply this correction but instead adopted for all stars the mean distance carefully derived by \citet{Baumgardt+2021}, to avoid enlarging the scatter in the subgiant branch of the cluster.
We used the mean absorption values per cluster from Table~\ref{tab:clusterdata}, and we applied the zero-point corrections and empirical calibrations to derive an extinction-corrected colour-magnitude diagram.

We then select MSTO and SGB stars of the clusters in the same way as done for the field stars in Sec.~\ref{sec:MSTOSGB} and Figure~\ref{fig:MSTOSGB}.
For NGC~6397 we additionally set a limit on bright magnitudes ($M_G>2$) to exclude horizontal branch stars.
This gives a selection of 4,374 stars in 14 clusters.

\subsection{Results per cluster and literature comparison}\label{sec:clusterres}

\begin{figure*}[htp]
\centerline{\includegraphics[width=0.99\textwidth]{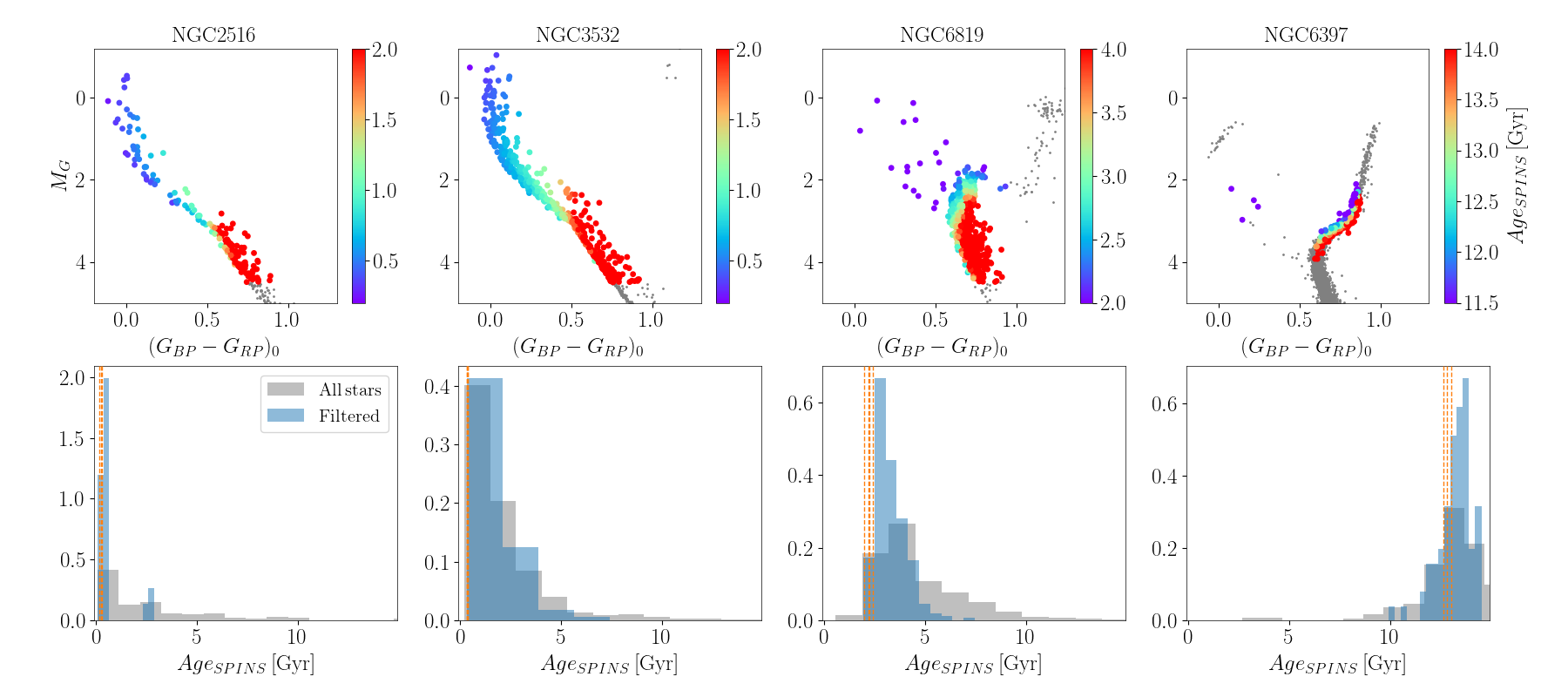}}
\caption{SPInS results for three of the analysed clusters: NGC 2516 (0.2 Gyr), NGC 3532 (0.4 Gyr), NGC 6819 (2.2 Gyr) and NGC 6397 (12.6 Gyr).
Top: intrinsic CMDs of the member stars (gray) and the selection of stars analysed by SPInS coloured by the obtained age (from the median value of the PDFs).
Bottom: age histograms of all analysed stars in gray; and a filtered subsample is shown in blue, selected as: relative uncertainties in age better than 20\% (and better than 500 Myr), excluding blue stragglers (BSS), and excluding stars in the equal-mass binary sequence.
Literature ages are marked with vertical dashed orange lines.}
\label{fig:resultsclusters_ex}
\end{figure*}

\begin{figure*}[htp]
\centerline{\includegraphics[width=\textwidth]{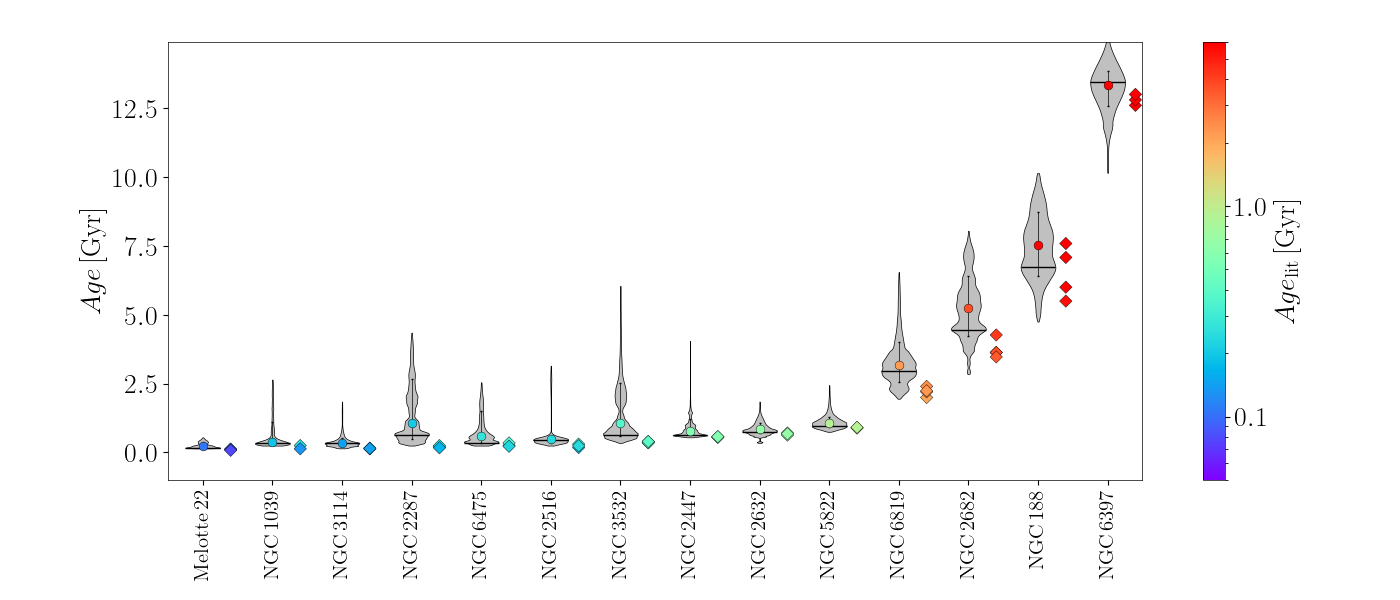}}
\caption{Comparison between the age quoted in the literature from studies indicated in Table~\ref{tab:clusterdata} (diamond symbols), and the age distribution per cluster obtained with SPInS in the form of violin plots. The horizontal black line depicts the mode of the age distribution. We also plot with filled circles and errorbars the age estimate coming from the median age per star instead of the full PDF.}
\label{fig:resultsclusters}
\end{figure*}

We run SPInS star by star with the observational constrains ($M_G$,(\bprp)$_0$) and metallicities from the literature as indicated in the previous subsection.
Since we use solar-scaled mixture evolutionary models, for the case of the globular cluster NGC 6397 we take into account its $\alphFe$ determination to re-scale the metallicity to mimic the $\alpha$ element enrichment, as explained in Sect.~\ref{sec:basti}.

The results from SPInS give a PDF sampled by the MCMC of the three fitted parameters (age, metallicity, mass), for which we can take the median value as the best estimate, and the 16{\th} and 84{\th} percentile to compute the uncertainties.
We show in Figure~\ref{fig:resultsclusters_ex} the results of the age determination star by star for four of the analysed clusters.
We plot the intrinsic CMDs per cluster, where each analysed star is coloured according to its median age.
We can extract several conclusions on the performance of our age determination method which we describe below.

\begin{itemize}
    \item We notice that stars in the lower main sequence (far from the turnoff) tend to have overestimated ages. This is no surprise since, for a given mass, an evolutionary track stays nearly in the same place of the main sequence for a long time.
These stars will therefore present broader PDFs in the age dimension, which, overall, will result in larger median ages, particularly for the less massive members of young clusters.
These lower main-sequence stars will also have larger uncertainties derived from the 16{\th}/84{\th} percentiles of the PDF, which means that we can filter them out using a cut on their relative uncertainties.

\item We can also see the biases due to the presence of blue stragglers (BSS) in old clusters such as NGC 6819 and NGC 6397, which will tend to provide younger ages than the real cluster age.

\item Finally, we also notice that  unresolved binaries (particularly those which belong to the equal-mass binary sequence) tend to provide older ages than a single star at a given magnitude, because its locus in the CMD corresponds well with an old turnoff star.
We have not done any hard cut on RUWE in this sample of stars because it penalised most of the stars belonging to the turnoff of few clusters (particularly NGC~2287 and NGC~2447), thus biasing the final results.

\end{itemize}

The discrepancies in the cases of BSS and equal-mass binaries are expected simply because we cannot use standard evolutionary models to describe multiple stellar systems and exotic objects.
In this test case using clusters, we take advantage of the fact that we can easily identify these stars by eye in a colour-magnitude diagram and remove them manually from the sample.
From the total of 4,374 stars in our sample of clusters, we identify 58 BSS and 464 binaries, which represents 11\% of the sample.
We have not found a way to filter them out in a general way for the entire range of possible ages in a scientific case of field stars.
However, the fraction of BSS and unresolved binaries found in the clusters allows us to set an order of magnitude of the "contamination" rate that one can expect from a sample of field stars selected in the same way.

In the bottom plots of Figure~\ref{fig:resultsclusters_ex} we show the histogram of the median age obtained per star in grey, and in blue we only represent stars filtered according to the following criteria:
(i) we remove stars which give large uncertainties on age, restricting the set of stars to those with relative uncertainties lower than 20\% and uncertainties better than 500 Myr (a cut on absolute age uncertainty is needed to filter old stars with large uncertainties), (ii) we exclude the identified BSS and binaries.
We notice a general improvement in the consistency with literature ages by applying these filters.
In Figure~\ref{fig:resultsclusters} we plot the median age values per cluster obtained from the blue histograms, with errorbars representing the 16{\th}/84{\th} percentiles, coloured by the mean literature age.

To have a better description of the involved uncertainties in a Bayesian manner we can describe the age PDF of the cluster as the sum of the individual PDFs of the member stars.
This allows us to keep the full information given by SPInS in the final age distribution, including stars with multiple solutions.
We show the age distribution for each cluster in Figure~\ref{fig:resultsclusters} representing in the form of violin plots the 95\% of the cumulative distribution functions. 
As was done for the previous case, we have filtered the stars previously identified as BSS, unresolved equal-mass binaries or stars with uncertainties larger than 20\% or 500 Myr.
We also show in diamond symbols the literature values for each cluster listed in Table~\ref{tab:clusterdata}.
In this way we are able to see the details of the posterior distributions, which are highly skewed, sometimes presenting multiple peaks.

In general, we see a good consistency between our age estimates and the literature ages.
For young clusters ($<1$ Gyr), even though the peaks of the distributions are very close to the literature values, we have a tendency to slightly overestimate the ages of these clusters due to a tail towards old ages.
These tails are the consequence of the uniform cut in absolute magnitude performed for our sample selection for all clusters.
A cut at $M_G<4.5$ makes us include a larger proportion of stars in the main sequence for young clusters compared to old clusters (see for instance NGC~3532 in Fig.~\ref{fig:resultsclusters_ex}).
As explained above, ages for young main sequence stars are in general overestimated because of their broad PDFs.
This overestimate is larger when taking a simple median of the stars per cluster, instead of combining up the individual PDFs (violin plots), where there is a clear peak of the PDF distribution very close to the literature values.
For clusters older than $\sim$500 Myr these tails are smaller because there is a larger proportion of stars in the turnoff, which dominate the cluster's PDF.
For the oldest case, the globular cluster NGC~6397, we find that the peak of our distribution is in nice agreement with literature estimates, which are very consistent among them, though a clear gradient of age across the subgiant branch is seen for this cluster in Fig.~\ref{fig:resultsclusters_ex}.

For intermediate-age clusters we tend to obtain relatively broad distributions, particularly for NGC 188 (for which literature ages are also quite diverse), but also NGC~2682 and NGC~6819.
These three clusters are also more distant than the sample of field stars analysed in Sect.~\ref{sec:results}, which is essentially limited to 1 kpc.
Thus, we expect larger uncertainties in photometry, manifested in the wider main sequences, which unavoidably give broader age distributions.

Particularly for these three clusters we find a large proportion of turnoff stars with double peaks in the PDFs, see for instance Fig.~\ref{fig:triangle}.
We have developed an algorithm to automatically detect the cases where the PDF shows multi-peaks (see Sect.~\ref{sec:appendix_MS} for a detailed explanation), which we have run on the full sample of cluster stars.
We have found that multi-peak solutions are in general found for stars placed near the turnoff loop.
This feature causes the stellar tracks to overlap particularly for tracks of mass $>1M_{\odot}$ at Solar metallicities.
In general, a filter in the overall age uncertainty removes a large fraction of multi-peak stars, but those for which the two peaks are relatively close are not removed.

\begin{figure}[htp]
\centerline{\includegraphics[width=0.5\textwidth]{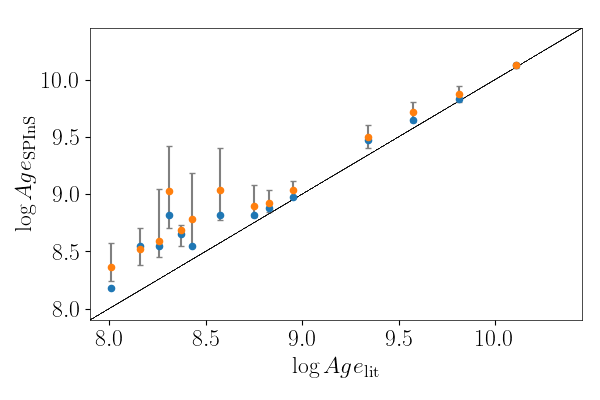}}
\caption{Comparison between the age quoted in the literature from studies indicated in Table~\ref{tab:clusterdata}, and the determination from SPInS using the median values of the filtered stars (orange), and the mode (blue).}
\label{fig:resultsclusters_summary}
\end{figure}

Overall, we find a good agreement of the cluster ages results' from SPInS when compared to literature results.
When we look at the comparison over the full age range in Fig.~\ref{fig:resultsclusters_summary}, we see a small tendency from our study to overestimate the ages with respect to literature, particularly for clusters younger than 1 Gyr, with a mean deviation of 200-500 Myr, depending on whether we use the mode or the median of the distribution for each cluster.

\section{Results: field stars}\label{sec:results}

\begin{figure}[htp]
\centerline{\includegraphics[width=0.5\textwidth]{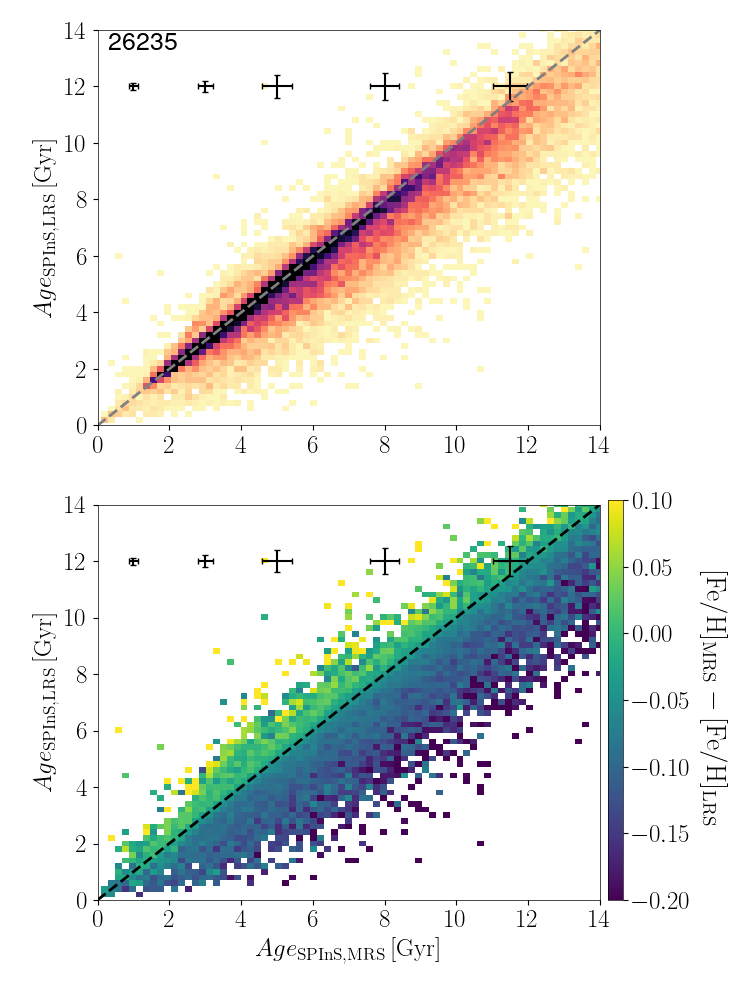}}
\caption{Comparison of the ages from the MRS-LASP and LRS samples obtained by SPInS. The number of stars is indicated in the plot, and median quoted errors per age bin are plotted. Top plot is coloured by density, and bottom plot by the difference in metallicity.}
\label{fig:comparisonsamples}
\end{figure}

We use SPInS to determine masses and ages of the selected MSTO and SGB stars with the configuration as explained in Sec.~\ref{sec:method}.
We use as observational constraints: absolute magnitudes from \gaiadr\ ($M_G$,(\bprp)$_0$), and metallicity from LAMOST.
Uncertainties in these parameters are assumed to follow Gaussian distributions.

As for the case of the globular cluster in Sect.~\ref{sec:clustersel}, we re-scale all the metallicities using the $\alphFe$ provided by LAMOST to mimic the $\alpha$ element enrichment, to be able to use solar-scaled evolutionary tracks (see Sec.\ref{sec:clusterres}).
For the MRS we compute two sets of ages, one using $\alphFe$ and $\feh$ from the main LAMOST pipeline (LASP), and another one using the abundances from LAMOST DR8 obtained using the label-transfer method based on a convolutional neural network (CNN), trained on APOGEE.
The uncertainty propagation of the scaled metallicities using the \citet{Salaris+1993} formula is not straightforward because it is likely that the uncertainties in the $\alphFe$ and $\feh$ of LAMOST are correlated.
Since LAMOST does not provide a specific correlation matrix between the two values, here we have propagated the uncertainties analytically as if the two quantities were uncorrelated, i.e. the quadratic sum of the $\alphFe$ and $\feh$ multiplied by the partial derivatives of the \citet{Salaris+1993} formula.
The partial derivative with respect to $\feh$ is always 1, and the one with respect to $\alphFe$ is always smaller than 1 taking into account the values of $\alphFe$ in our sample.
We have decided to assume a value of exactly 1 for the latest derivative in order to slightly inflate the final errors, and possibly mitigate the fact that we are assuming independent measurements for $\alphFe$ and $\feh$.

We provide age estimates for 35,096 and 243,768 stars based on the metallicities and $\alpha$ element abundances of the LAMOST DR8 MRS and LRS samples, respectively.
For the case of the CNN values of MRS, the sample includes 34,779 stars with valid abundances from CNN.

In Figure~\ref{fig:comparisonsamples} we show the comparison of the ages derived for the MRS-LASP and the LRS for the stars in common, which shows an overall good agreement.
We notice that, even though the peak of the distribution stays in the 1:1 line, the overall shape is slightly asymmetric, in the sense that LRS ages seem slightly underestimated with respect to MRS.
We have investigated the possible causes of this and we find that there is a small systematic overestimate in the metallicity of the LRS sample with respect to the MRS one ($\sim0.06$ dex in mean), which could explain the small difference in age.

The comparison between MRS-LASP and MRS-CNN also shows a slight tendency of ages derived from CNN to be underestimated with respect to LASP (mean value of -271 Myr) particularly for the metal-poor stars.
This difference can be correlated with the difference between the CNN and LASP metallicites (see Fig.~\ref{fig:LASPCNN_diffage}), and with a difference in the estimation of the atmospheric parameters from the two methods.
As explained in Sect.~\ref{sec:data}, for our age determination, we choose not to use spectroscopically derived atmospheric parameters, but only de-reddened photometry and spectroscopic metallicity.
However, SPInS provides posteriors in all parameters of the input grid, including $\teff$, which we compare to spectroscopically derived ones in Fig.~\ref{fig:LASPCNN_Teff}.
The figure shows that LASP estimates seem to be more coherent with SPInS determinations than CNN ones, which have an overall offset of 124 K and a trend towards hot stars.

\begin{figure*}[htp]
\centerline{\includegraphics[width=\textwidth]{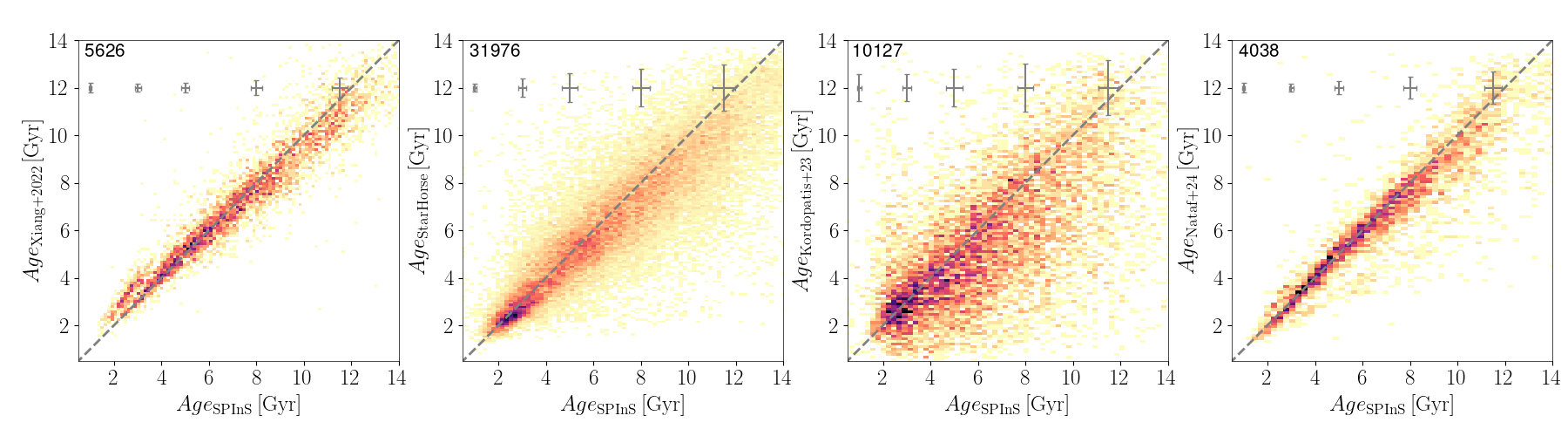}}
\centerline{\includegraphics[width=\textwidth]{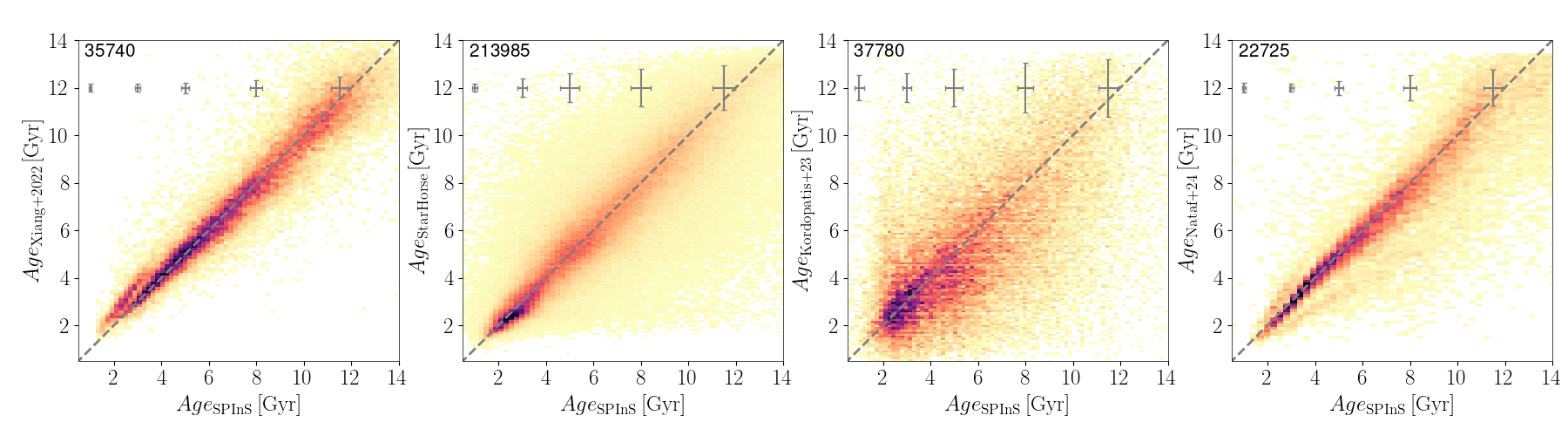}}
\caption{Comparison of the ages from the MRS (top) and LRS (bottom) samples obtained by SPInS with literature estimates from \citet{Xiang+2022}, {\tt StarHorse} obtained in \citep{Queiroz+2023}, \citet{Kordopatis+2023} and \citet{Nataf+2024} (from left to right). The number of stars is indicated in each plot, and median quoted errors per age  bin are plotted.}
\label{fig:comparisonlit}
\end{figure*}

\subsection{Comparison with large age catalogues}

We compare the resulting median age per star obtained by SPInS of the MRS and LRS samples with ages coming from recent literature catalogues for stars in common in Figure~\ref{fig:comparisonlit}.

The catalogue of \citet{Xiang+2022} contains 247k stars, and they used a similar method as in this study to derive ages and masses from LAMOST DR7 using Yonsei-Yale stellar isochrones \citep{Demarque+2004}.
The main difference between their approach and ours is the fact they use as inputs spectroscopic estimates of the absolute magnitude in the $K$ band, instead of \emph{Gaia} photometry, and estimates of $\teff$ inferred by the data-driven method run in LAMOST DR7 (DD-Payne).
The catalogue of \citep{Queiroz+2023} using {\tt StarHorse} for LAMOST DR7 MRS/LRS sub-giant MSTO stars, and contains age estimates as well as other stellar astrophysical parameters for 120k and 1.3M stars, respectively.
{\tt StarHorse} is a Bayesian isochrone-fitting code that uses as many input observables as available (including parallaxes, spectroscopic stellar parameters, and multi-band photometry) to compute the likelihood of the observed quantities for a grid of PARSEC 1.2S + COLIBRI stellar models \citep{Marigo+2017}.
Aside from the canonical priors, like the IMF, {\tt StarHorse} uses a prior on the interstellar extinction based on a 3D extinction map as well as generous space density, age, and metallicity priors for the Galactic discs, bulge, halo, GC system, and nearby dwarf galaxies.
The catalogue obtained by \citet{Kordopatis+2023} contains 5 million stars with stellar ages computed in four different ways, depending on different combinations of parameters to project on the PARSEC 1.2S + COLIBRI isochrones.
Here we used their final "optimal" ages, which they provide as a combination of the four different projections (see their section 2.6.2).
Following the recommendations in \citet{Kordopatis+2023} we filter stars whose estimated relative age uncertainty is larger than 50\%.
Finally, the recent catalogue by \citet{Nataf+2024} provides ages for 289k stars obtained from multi-band photometry with 3D extinction maps and without any information of the metallicity.

We find a general good agreement in the ages from SPInS with the aforementioned catalogues from the literature.
The comparison with \citet{Xiang+2022} shows a marked difference for stars between $\sim2-4$ Gyrs where our results are in general 500 Myr younger than those of \citet{Xiang+2022}.
These are in general hot stars ($\teff>6000$) for which we noticed that the effective temperatures derived with LAMOST (DR8) LASP pipeline and the data driven approaches DD-Payne/CNN differ systematically, in the sense that DD-Payne always finds a cooler $\teff$ for hot stars.
This bias could tentatively explain the observed difference, since a cooler $\teff$ for a given absolute magnitude and metallicity will generally provide older ages.
However, this is an indirect explanation because SPInS ages do not use LASP $\teff$ but only information from derredenned photometry.
It is also possible that this difference comes from the two different sets of stellar evolutionary models used.
Overall though, we obtain a mean difference and scatter of $-0.06/1.12$ Gyr with \citet{Xiang+2022} for the LRS.
The comparison with \citet{Queiroz+2023} is the one that provides the most stars in common in the two samples and shows a good agreement, with a mean difference and scatter of $-0.28/1.87$ Gyr.
In particular it does not show the previously mentioned artefact for young ages.
The comparison with \citet{Kordopatis+2023} is more scattered and shows horizontal stripes, probably due to the mild pixelization of the GSP-Spec results \citep{RecioBlanco+2023}, or alternatively to a different type of interpolation along stellar tracks.
The mean difference and scatter are $-0.55/2.63$ Gyr, though the scatter decreases to 2.2 Gyr if we only keep stars with GSPSpec flags equal to 0 (except for noise, allowed to be 0 or 1).
The larger scatter with this catalogue is well explained by the difference in the metallicity between the two catalogues (see Fig.~\ref{fig:comp_met}), highlighting the importance of the quality of the metallicity in the age estimates.
The comparison with \citet{Nataf+2024} shows a very good agreement with their ages, with only faint structures outside the 1:1 line for the case of the youngest stars, which are visible in the LRS sample.
We highlight, however, that their estimations of the metallicity have a significant offset of $\sim0.15$ dex with respect to LAMOST determinations (see Fig.~\ref{fig:comp_met}), as they already cite in their original paper.
Surprisingly, this does not seem to have a big effect in the comparison with their ages, unlike the case of \citet{Kordopatis+2023}.
\citet{Xiang+2022} and \citet{Queiroz+2023} metallicities are very similar to ours, as seen in Fig.~\ref{fig:comp_met}, because they come from LAMOST even though previous data releases.

In general, we find very little systematics between the literature datasets and the results of SPInS, except for the youngest stars present in the \citet{Xiang+2022} dataset.
We expect then similar offsets in these literature catalogues as those found in the present study using clusters (Sec.~\ref{sec:clusterres}, Table~\ref{tab:clusterdata}).

\subsection{Kinematics}\label{sec:kinematics}
We have used the 6D phase space information provided by \gaiadr\ with the radial velocities provided by the LAMOST DR8 catalogues (MRS and LRS) to compute action angle variables and orbital parameters of the two samples with the \texttt{galpy} package \citep{Bovy2015}.
Galactocentric cartesian positions ($X,Y,Z$) and cylindrical positions and velocities ($R,\phi,Z,V_R,V_{\phi},V_Z$) were computed assuming the Solar values: $(X,Y,Z)_{\odot} = (-8.34,0,0.027)$ kpc from the center of the Galaxy, an azimuthal velocity at the solar radius of $240\,\kms$, and the velocity of the Sun with respect to the local standard of rest (LSR) as $(-11.1, 12.24, 7.25)\,\kms$ \citep{schonrich10}.
Uncertainties were computed using the quoted individual uncertainties and covariance matrix with the \texttt{galpy} utils package.

Action-angle variables and orbital parameters were obtained using the axisymmetric potential MWPotential2014 implemented in \texttt{galpy}.
The integration of the orbits was done up to 1 Gyr with 1000 steps, to obtain eccentricity, apocentre, pericentre, maximum distance from the Galactic plane $Z_{max}$, and guiding radius $R_g$.
The obtained kinematical parameters will be used in the next subsections.

\begin{figure*}[htp]
\includegraphics[width=0.33\textwidth]{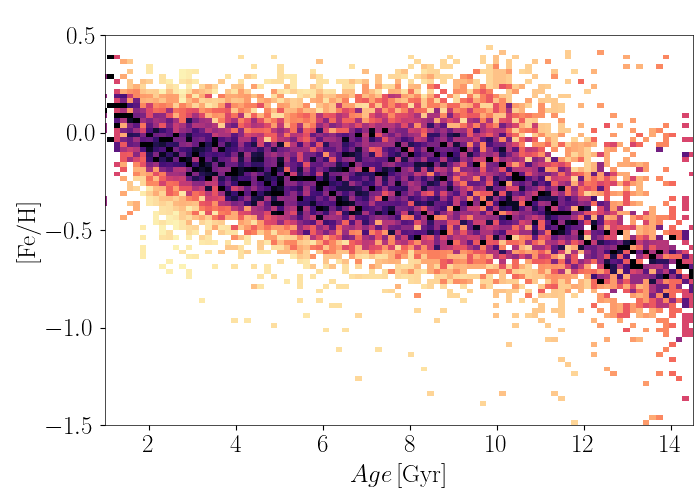}
\includegraphics[width=0.33\textwidth]{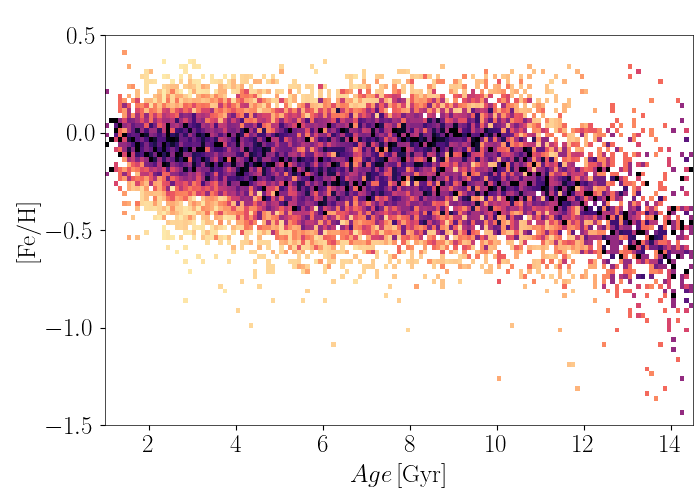}
\includegraphics[width=0.33\textwidth]{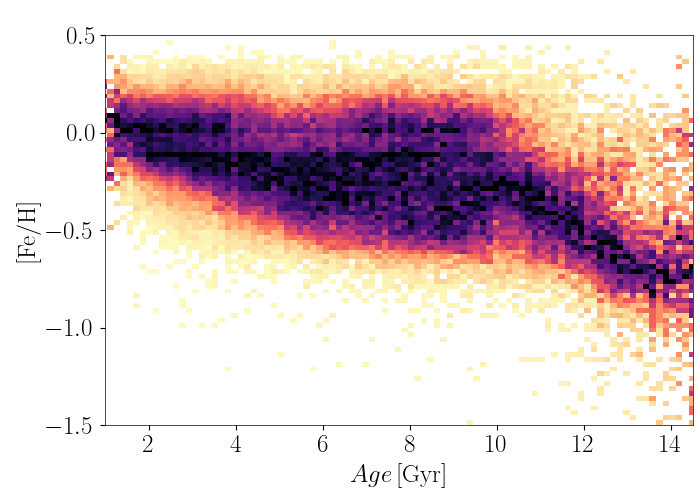}
\caption{Age-metallicity relation obtained from the MRS (left using LASP abundances, and middle using CNN abundances), and LRS (right) samples. The colour corresponds to the normalised density of stars at each age bin.}
\label{fig:AMR}
\end{figure*}

\section{Discussion: age-abundance-kinematics relations}\label{sec:discussion}
We have used our age estimates for the MRS and LRS samples to investigate the age-$\feh$\ and age-$\alphFe$ relation in the Solar neighbourhood.
We use the orbital parameters computed in Sect.~\ref{sec:kinematics} to dissect the age-abundance relations.

Following the validation tests performed on open and globular clusters in Sect.~\ref{sec:val}, we have applied a filter to the age catalogue only keeping stars with $\delta$Age/Age$<0.2$ and $\delta$Age$<500$ Myr.
These filters reduce the size of the two samples to 20,822 stars in the MRS (19,911 for the MRS-CNN) and 132,211 in the LRS.
They particularly remove a large fraction of stars at ages $<1$ Gyr, and a group of stars with solar metallicities with very old inferred ages (most probably low main sequence stars).
We also filter the stars which give multiple solutions in SPInS, detected as explained in the Sec.~\ref{sec:appendix_MS}, which remove 3,589 and 24,490 additional stars in the MRS and LRS samples, respectively.

\subsection{Age-metallicity relation}

\begin{figure*}[htp]
\includegraphics[width=0.49\textwidth]{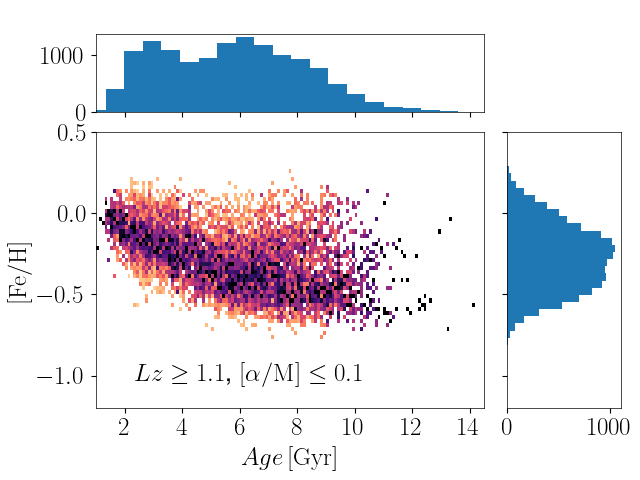}
\includegraphics[width=0.49\textwidth]{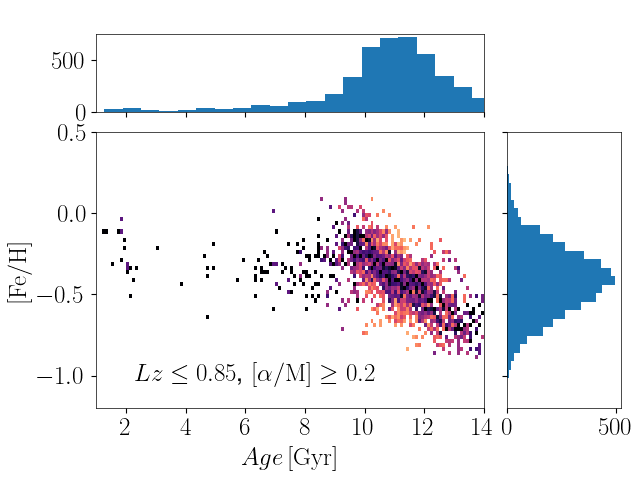}
\caption{Age-$\feh$\ relation obtained with the LRS sample, isolating the $\alphFe$-enhanced thin (left with 16,800 stars) and thick (right with 7,623 stars) disks. The colour corresponds to the normalised density of stars at each age bin (only bins with more than 1 star are shown).}
\label{fig:AMR_thinthick}
\end{figure*}

We plot the age-$\feh$\ relation (AMR) of our samples in Fig.~\ref{fig:AMR} coloured by the stellar density.
The histograms have been normalised for each age bin by its maximum value.
The plots for the MRS and the LRS show a hint of two tight sequences, overall consistent with the already known bi-modality of the AMR \citep{Nissen+2020,Jofre2021,Xiang+2022,Anders+2023}, see also discussions in \cite{Sahlholdt+2022} and Cerqui et al (in preparation).
For the LRS we see a horizontal stripe  at $\feh$\ between 0 and -0.1 which causes an underdensity of stars in this region.
We have not found the cause of this, but it is most probably an artefact or a bias in the LAMOST LRS abundance pipeline.
For the MRS the overall picture is similar when we use the ages obtained using the CNN (middle panel) and LASP abundances (left panel).
The thin disk sequence seems more prominent when using LASP abundances.

In Fig.~\ref{fig:AMR_thinthick} we dissect the AMR of the LRS using simultaneous cuts in angular momentum and $\alphFe$ abundances, as done in \citet{Xiang+2022}.
A low $\alphFe$\ branch corresponding to the chemical thin disk is clearly visible up to $\sim$6 Gyr for stars with $L_z\geq1.1$\footnote{We give $L_z$ in units of $[\mathrm{8.34kpc\times240\kms}]$} and $\alphFe\leq0.1$.
The sequence seems to extend up to 9 Gyr but shows a tail of stars at higher metallicities for ages older than 6 Gyr, which possibly correspond to a population of radial migrators.
On the other hand, selecting stars with $L_z\leq0.85$ and $\alphFe\geq0.2$ yields a tight sequence starting at $\sim$9 Gyr corresponding to the $\alphFe$-enhanced disk, or thick disk, which has a steeper slope in the AMR.
Similar selections also isolate the  two sequences in the MRS sample, but with a much smaller number of stars.
Between 6 and 9 Gyr, particularly when selecting  the intermediate cuts with $L_z$ and $\alphFe$, we observe a mixture of the two populations, without a clear split of the two sequences.

The age distribution of the low $\alpha$ disk shows two distinct peaks at 3 and 6 Gyr with a lower density "gap" around 4 Gyr.
We have checked that the gap is still present when we use the sample without filtering the multiple solution stars, to ensure that it is not an artifact of removing stars in a particular region of the CMD.
Similar gaps in the young disk are observed in other recent samples of ages, but they are not always located at the same age \citep[e.g.][]{Sahlholdt+2022,Anders+2023,Queiroz+2023}.
There is a clear relation between age and metallicity in the young peak, and then a more blurred dependency is seen in the second peak after the gap, with a hint of a possible change of the slope.
The multimodal age distribution coupled with the tight metallicity dependence also imprints two very close peaks in the metallicity distribution, seen in the projected histogram.
We do not find in our AMR the V shape found by \citet{Xiang+2022}, who showed an additional branch with a positive slope.
For the high $\alpha$ disk we obtained a more restricted age distribution with a unique peak at around 11 Gyr, and a long asymmetric tail towards younger ages, which could be due to young $\alpha$ rich stars \citep[e.g.][]{Cerqui+2023}.
To better understand these dependencies an additional and careful analysis of the selection function is needed.

\begin{figure}[htp]
\includegraphics[width=0.49\textwidth]{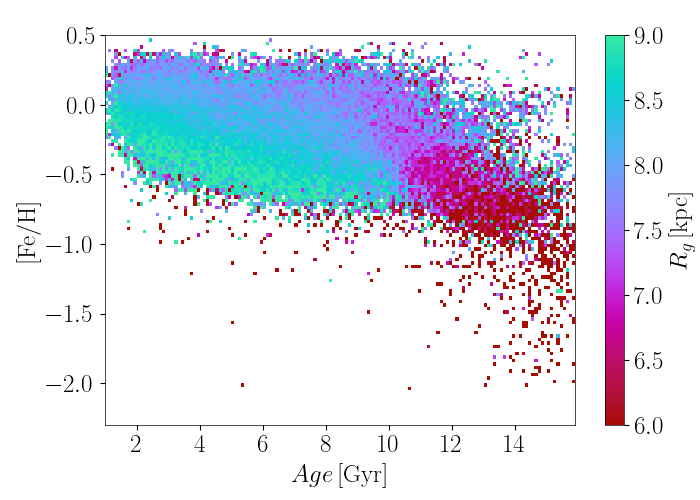}
\includegraphics[width=0.49\textwidth]{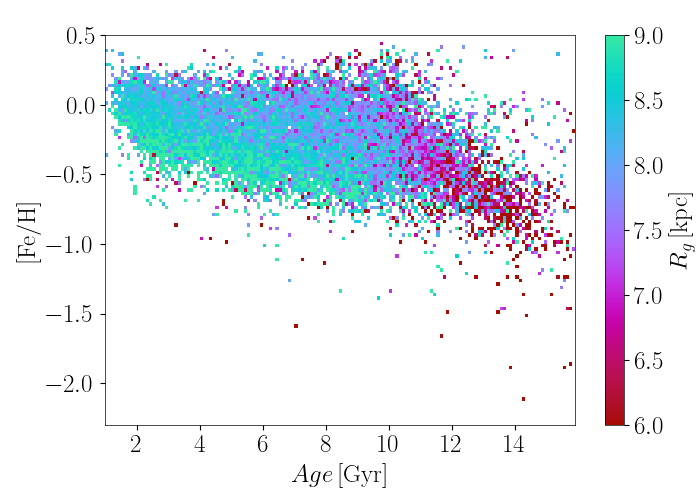}
\caption{Non-normalised AMR for the LRS (top) and MRS-LASP (bottom) coloured by the mean guiding radius per bin.}
\label{fig:AMR_Nnorm}
\end{figure}

We plot in Fig.~\ref{fig:AMR_Nnorm} a 2D histogram of the AMR coloured by the mean guiding radius ($R_g$) per bin for the full LRS and MRS-LASP samples (no filter in $L_z$ or $\alphFe$).
The non-normalised AMR highlights better the few stars in our sample with metallicities lower than -1, which form a vertical structure particularly visible in the LRS, without any clear dependence in age, and which present small $R_g$.
We also see a dependence of the mean guiding radius along the high $\alpha$ disk sequence.
Overall, the dependence of the AMR with guiding radius is consistent with the general understanding of the Galactic radial metallicity profile coupled with radial mixing \citep[e.g.][]{Haywood2008, Schonrich+2009,Minchev+2014,Hayden+2015,Frankel+2018,Hayden+2020,Haywood+2024}.
Even though our sample is made of stars in the very local Solar neighbourhood, we can see the effect of radial migration which, as time goes by, brings stars born at different Galactocentric radii to the Solar radius.
Thus, at a given age we see a clear gradient in guiding radius for the young ($<10$ Gyr) AMR region, where low metallicity stars have a larger guiding radius than high metallicity stars.

\begin{figure*}[htp]
\includegraphics[width=\textwidth]{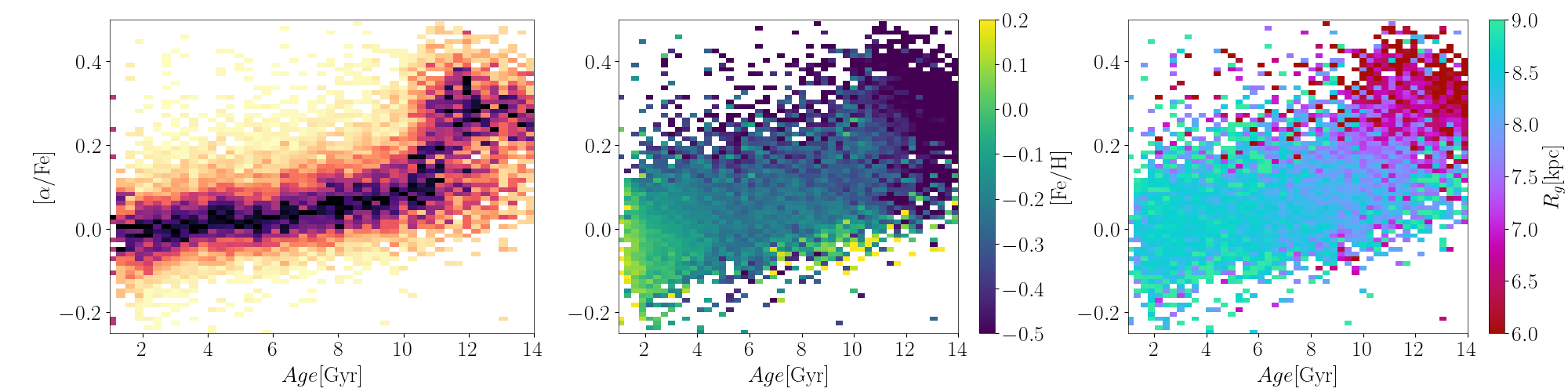}
\includegraphics[width=\textwidth]{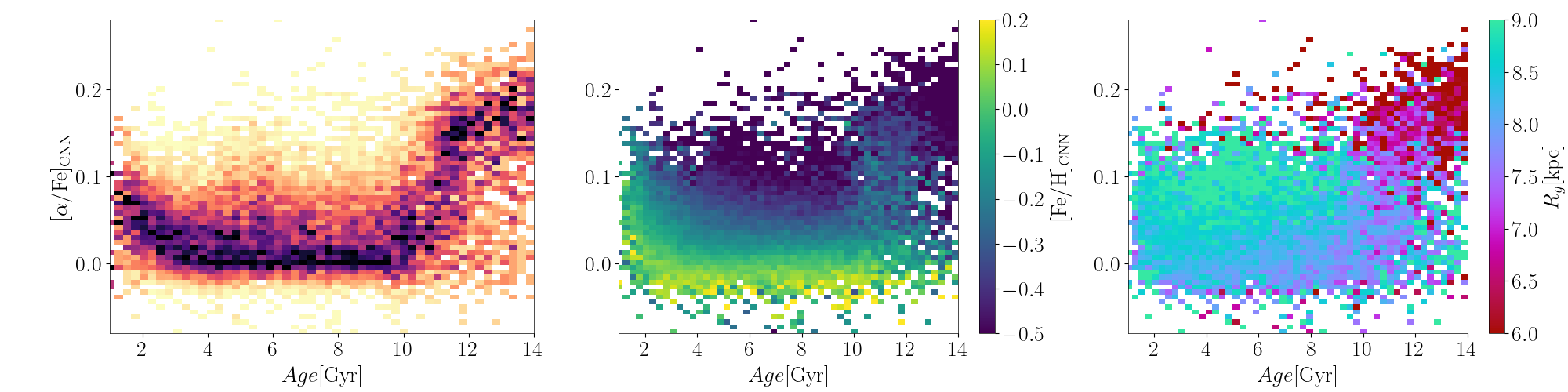}
\caption{$\alphFe$ vs age relation obtained for the MRS sample using LASP abundances (top), and CNN abundances (bottom). In the left panel, the colour corresponds to the normalised density of stars per age bin, in the middle and right panel we colour each bin by the mean value of $\feh$ and $R_g$ per bin values, respectively.}
\label{fig:alphaAge}
\end{figure*}

\subsection{$\alphFe$-age relation}

We plot the relation between $\alphFe$ and age in Fig.~\ref{fig:alphaAge} for the MRS.
In the left panels the colour represents the stellar density, and in the middle and right panels we plot a 2D histogram weighted by the mean $\feh$ value per bin and the mean $R_g$ per bin, respectively.
We show the relation when using the $\feh$ and $\alphFe$ from LASP abundances in the top plots, and in the bottom plots using the CNN values (notice the change of vertical scale).
The usage of the two sets of abundances shows significant differences in the $\alphFe$-age relations. 
In the case of LASP we see a slow increase of the $\alphFe$-age relation for stars younger than $\sim 9-10$ Gyr.
A high-$\alpha$ sequence with a much higher slope starts to be visible at $10$ Gyr, which is usually linked to the chemically defined thick disk.
In the case of CNN abundances, the thin disk shows a much flatter relation up to $\sim 9-10$ Gyr, with a much steeper increase of the $\alphFe$ for the thick disk sequence.

We notice a blob of stars with old ages ($>12$ Gyr) which seem to have constant and even decreasing $\alphFe$, and which gives the impression of a decreasing $\alphFe$ vs age sequence for the oldest stars in the LASP case.
In the case of CNN, this blob stays at roughly constant values of $\alphFe$, as the end of the thick disk.
This feature corresponds essentially to a group of stars at relatively low metallicities ($\feh<-0.6$)  which, in the LASP case, seem to have decreasing $\alphFe$ abundances in the $\feh$ vs $\alphFe$ plane.
This feature is also seen in other age samples using LAMOST abundances such as \citet{Queiroz+2023}.
In fact, in this region there are few stars, but normalising the histogram at each age bin enhances this feature.
We have not found a clear reason for this, it is possible that the values of the overall $\alpha$ abundances for low metallicity stars in LAMOST suffer from some biases due to the intrinsic difficulty of deriving abundances at low metallicities.

In the middle panels of Fig.~\ref{fig:alphaAge}, we see the imprint of the AMR (Fig.~\ref{fig:AMR}) as a colour gradient along the young chemically defined thin disk sequence.
In the case of the CNN abundances we see a more clear stratification of the metallicities across the thin disk.
The plot when using the LASP abundances shows a significant group of intermediate age stars (7-10 Gyr), with high $\feh$\ ($>0.1$) and low $\alphFe$ abundances.
In the right hand plot we see how these stars tend to have slightly smaller guiding radii with respect to the rest of stars at the same age.
With a mean $R_g\sim6.7$ kpc, most of these stars are most likely to be radial migrators coming from the inner disk.
This group of stars is not clearly identified when using the CNN abundances.
However, in the case of the CNN abundances, we see a more clear stratification of the thin disk with guiding radius (bottom right plot), showing the effects of radial migration to the $\alphFe$ vs age sequence.

Overall, there are remarkable differences between the LASP and the CNN values in the $\alphFe$-age relation, which can lead to different conclusions on the enrichment history of the thick and thin disks.
The difference comes mainly from the $\alphFe$ values, which have a very poor correlation among the two different determinations.
This is a clear example of the importance of the precision and accuracy of the atmospheric parameters and chemical abundances obtained from spectroscopic surveys across the full metallicity range.

\subsection{Age-velocity dispersion relation}

\begin{figure}[htp]
\includegraphics[width=0.5\textwidth]{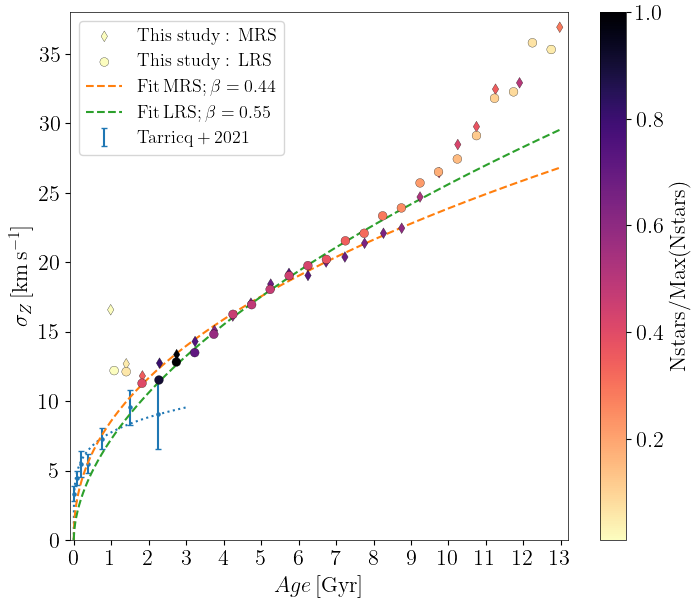}
\caption{Age-velocity relation for the MRS and LRS samples coloured by the relative number of stars in each point (divided by the maximum).
Power law fits to the $2<\mathrm{Age}<9$ Gyr range are plotted in orange and green dashed lines.
A comparison with the AVR obtained for clusters in the solar neighbourhood \citep{Tarricq+2021} is shown in blue dots and their power law fit in blue dashed line.}
\label{fig:AVR}
\end{figure}

The relation between the dispersion in the vertical velocity $v_z$ and age in the Solar neighbourhood is often called the age-velocity relation (AVR), and has been studied for decades using different tracers \citep{Wielen1985,Nordstrom+2004,Casagrande+2011}.
It is generally described as a power law, though its slope is debated.

We have computed the dispersion in $v_z$ in $\sim25$ bins of age for the filtered MRS and LRS.
We have additionally applied a cut on the uncertainty of $v_z$ of $<2\,\kms$ to avoid outliers.
The width of the bins are around 500 Myr, but were slightly tuned for the MRS for old ages to avoid having bins with very few stars.
We plot the age-velocity relation (AVR) obtained for the two samples in Fig.~\ref{fig:AVR} coloured by the number of stars, relative to the bin with the maximum number.
We have computed the uncertainties in $\sigma_z$ sampling the uncertainties of the cylindrical velocities 1,000 times.
Obtained uncertainties are smaller than $<0.1 \kms$ in all bins, and are thus smaller than the point size.

We overplot the AVR obtained for a sample of open clusters in the Solar neighbourhood obtained by \citet{Tarricq+2021}, which cover the age range of [0,2.5] Gyr.
We overplot with a dotted blue line the power-law fit that they performed, with a slope value of $\beta_{\mathrm{OC}}=0.19\pm0.03$.

The two youngest age bins ($<$2 Gyr) of both the MRS and the LRS seem to indicate a flattening of the $\sigma_z$ at around 12 $\kms$ or even increase in the dispersion, behaviour which is not reproduced by the open cluster sample.
The statistics of these bins are smaller than the rest, particularly in the MRS, which count less than 100 stars.
Our age determination in this range of ages gives slightly overestimated values, as seen for the youngest clusters in Fig.~\ref{fig:resultsclusters}.
Additionally, taking into account the validation tests done in Sect.~\ref{sec:val}, it is also possible that the youngest bins have a contamination due to blue stragglers, which would then have typical kinematics of older stars, causing an increase in $\sigma_z$.
Overall, we consider that the two youngest bins are most probably not representative of the true AVR of the Galaxy.

As done by \citet{Anders+2023} we perform a power-law fit to our two samples in the age range $2<\mathrm{Age}<9$ Gyr, in orange and green, for the MRS and LRS respectively.
We limited the age range to avoid the two youngest age bins, and the oldest regions where we see a departure from a simple power-law.
The uncertainties of the obtained coefficients are extremely small, of the order of $10^{-4}$.
We see how the open cluster sample nicely matches the power law fitted to the field stars, and complements it in the younger age range.
The oldest point of the open cluster sample, though still compatible with our trend, has a remarkable lower value.
This decrease is probably related to the low statistics of the open cluster populations at ages $>2$ Gyr (reflected by the large uncertainty on the dispersion), which has been attributed to a higher destruction rate of old open clusters.
As a direct consequence, the obtained slope with the open clusters is smaller compared to the field stars.

On the old end, a departure from the power-law fit is highlighted by a steepening of the $\sigma_z$-age relation at 9-10 Gyr.
This has been found in the literature and has been attributed to disk stars which have been kinematically heated due to the Gaia-Sausage-Enceladus merger event \citep{DiMatteo+2019,Belokurov+2020}.


\section{Conclusions}\label{sec:conclusions}
Reliable stellar ages, coupled with 6D phase space information and chemical abundances, are essential to understand the evolution of the Milky Way.
In this paper, we exploit the possibility of using the absolute colour-magnitude diagram from \gaiadr\ photometry only, coupled with spectroscopic estimates of the $\feh$ and $\alphFe$ from the LAMOST DR8 medium and low resolution samples.

We use the public SPInS code \citep{Lebreton+2020} to obtain age estimates and reliable uncertainties of individual main-sequence turnoff stars and subgiant stars.
We implement new tools in the code, which are now public, to obtain an automatic evaluation of the convergence of the MCMC sampler (see Sect.~\ref{sec:mcmc}), and an option to automatically detect multiple peaks in the case of a multi-modal posterior distribution (see Sec.~\ref{sec:appendix_MS}).

We test our strategy with a sample of 4,374 stars in 14 stellar clusters including the old metal-poor globular cluster NGC~6397.
With this sample we are able to investigate the possible systematic and random uncertainties which can affect the age distribution of a sample of field stars selected with the same criteria and analysed with the same code.
On one side, we evaluate the effect of the presence of main sequence stars in the sample which biases the cluster ages by overestimating them.
These stars can be mostly removed by filtering our uncertainties to 20\% and with a hard cut at 500 Myr for the oldest clusters.
We also evaluate the effect of the presence of non-single stars, particularly blue stragglers and unresolved binaries, which, unsurprisingly under- and over-estimate the cluster ages, respectively.
Statistically in a sample of field stars selected in the same way, the non-single stars would represent around 11\% of the sample.

We see how the combination of the posterior distribution of the selected cluster stars yields age distributions with prominent peaks at the cluster's quoted literature ages, which include ages obtained from isochrone fitting, asteroseismology, eclipsing binaries and the white dwarf cooling sequences.
We apply the same strategy to an exquisite sample of field stars with very small photometric uncertainties (in general smaller than 0.04 in $M_G$, and of the order of 0.01 in $G_{BP}-G_{RP}$).
This sample has radial velocities, $\feh$ and $\alphFe$ abundances coming from LAMOST DR8 medium resolution MRS sample (35,096 stars which have LASP abundances and 34,779 stars which have CNN abundances) and/or low resolution LRS sample (243,768 stars).
By combining the obtained ages with the 6D phase space and the LAMOST chemical abundances, we are able to investigate the following points:

\begin{itemize}
    \item The age-metallicity relation (AMR) with the LRS and MRS samples, obtaining a consistent picture with its already known-bimodality.
    \item We dissect the AMR from the LRS with simple cuts in angular momentum and $\alphFe$ abundances, and we are able to isolate a low $\alphFe$ branch visible at young ages up to 6-9 Gyr, and a high-$\alphFe$ branch with a steeper slope in the AMR which covers ages older than 9 Gyr.
    \item While the high-$\alphFe$ branch shows a unique peak in the age distribution around 11 Gyr, the low-$\alphFe$ branch appears much more complex with at least two peaks at 3 and 6 Gyr. A better understanding of the selection effects is needed to correctly interpret this distribution.
    \item Using the MRS we study the $\alphFe$ vs age relation, which shows very different pictures when using LASP or CNN abundances. This highlights the importance of the precision and accuracy of chemical abundances from spectroscopic surveys. We investigate the features in the $\alphFe$-age relation coupling it with the $\feh$ and the guiding radius.
    \item The LASP $\alphFe$ abundances show a slowly increasing trend for stars younger than 10 Gyr and a high $\alpha$ sequence that starts to be visible at 10 Gyr. We detect a population of radial migrators form the inner disk which is present in this sample for ages between 7-10 Gyr.
    \item The CNN abundances show a much flatter slope of the thin disk stars, with a steeper thick disk sequence starting at 10 Gyr. A clear stratification of $\feh$ and $R_g$ is visible across the $\alphFe$ range for the thin disk.
    \item We investigate the age-velocity dispersion relation (AVR) in 25 age bins, showing a consistent power law relation in the range 2-9 Gyr with a slope of 0.44 and 0.55, respectively for the MRS and LRS samples.
    \item The simple power law breaks down in our sample for ages lower than 2 Gyr, though the two younger age bins contain fewer stars. Additionally, from the results of our tests using clusters, our method tends to overestimate the ages of younger stars, thus making this age range not reliable.
    \item Without these two youngest bins, the obtained AVR is compatible with that obtained with a sample of open clusters previously analysed in the literature.
    \item The AVR also breaks down at ages older than 9 Gyr, significantly increasing the vertical velocity dispersion, possibly related to the Gaia-Sausage-Enceladus merger event.
\end{itemize}


\begin{acknowledgements}
We warmly thank Rosine Lallement for her help and advice when using the extinction maps, and Santi Casissi for his advice when using the BaSTI stellar evolutionary tracks.

This work was partially supported by the Spanish MICIN/AEI/10.13039/501100011033 and by "ERDF A way of making Europe" by the “European Union” through grant PID2021-122842OB-C21, and the Institute of Cosmos Sciences University of Barcelona (ICCUB, Unidad de Excelencia ’Mar\'{\i}a de Maeztu’) through grant CEX2019-000918-M. FA acknowledges the grant RYC2021-031683-I funded by MCIN/AEI/10.13039/501100011033 and by the European Union NextGenerationEU/PRTR.

This work has made use of data from the European Space Agency (ESA) mission Gaia (\url{http://www.cosmos.esa.int/gaia}), processed by the Gaia Data Processing and Analysis Consortium (DPAC, \url{http://www.cosmos.esa.int/web/gaia/dpac/consortium}). We acknowledge the Gaia Project Scientist Support Team and the Gaia DPAC. Funding for the DPAC has been provided by national institutions, in particular, the institutions participating in the Gaia Multilateral Agreement.

Guoshoujing Telescope (the Large Sky Area Multi-Object Fiber Spectroscopic Telescope LAMOST) is a National Major Scientific Project built by the Chinese Academy of Sciences. Funding for the project has been provided by the National Development and Reform Commission. LAMOST is operated and managed by the National Astronomical Observatories, Chinese Academy of Sciences.

This research made extensive use of the SIMBAD database, and the VizieR catalogue access tool operated at the CDS, Strasbourg, France, and of NASA Astrophysics Data System Bibliographic Services.
\end{acknowledgements}


\bibliographystyle{aa} 
\bibliography{biblio} 


\appendix
\section{Additional tests and figures}\label{sec:appendix}

\subsection{Detecting multiple solutions in SPInS}\label{sec:appendix_MS}

In the following section we explain the details of the algorithm that detects multiple solutions in SPInS output, which has now been implemented in the public SPInS version.
This algorithm can be activated by the user, if needed, using the configuration file.

\subsubsection*{Implemented algorithm}
The algorithm's primary objective is to identify and isolate distinct solutions within multimodal distributions in stellar parameter inference, especially for datasets produced using the SPInS and AIMS \citep{Reese2016} programs. 
Hierarchical Density-Based Spatial Clustering of Applications with Noise \citep[HDBSCAN][]{Campello+2013} was chosen for its ability to find groups of varied densities without requiring a fixed number of groups, making it very versatile and appropriate for our requirements.
Input data for HDBSCAN in the case of SPInS include the model parameters of the MCMC samples, namely the initial mass $\log_{10}(M/M_\odot)$, metallicity [M/H], and Age (in Myrs).
These parameters are essential for the grouping process, while the log-likelihood values ($\text{ln} P$) are used after the grouping process to identify the optimal parameters for each group.
The data is standardized using StandardScaler, which removes the mean from the MCMC samples and scales them to unit variance.

HDBSCAN requires two primary parameters\footnote{\url{https://pberba.github.io/stats/2020/01/17/hdbscan/}} to be set: `min\_cluster\_size' and `min\_samples', as well as several hyperparameters such as `cluster\_selection\_method', a distance threshold `cluster\_selection\_epsilon', and the `allow\_single\_cluster' parameter which allows the detection of a single cluster.
After testing the algorithm in the specific case of SPInS results, we have chosen to use the `cluster\_selection\_method'=`leaf', `cluster\_selection\_epsilon'=0.3, `allow\_single\_cluster'=True, and `min\_cluster\_size' = `min\_samples'.
We have seen that the detection worked efficiently when setting `min\_samples' to 1\% or 2\% of the total number of samples, with a small dependence on the particular case.
Therefore the algorithm performs two runs with HDBSCAN using, respectively, the two values of `min\_samples', thus yielding two sets of labels which specify to which group each MCMC sample belongs.
The algorithm then compares the number of groups created by each run, and chooses the one that produces more groups.
If both runs generate the same non-zero number of groups, the sizes of the smallest groups are compared, and the run with the most samples in the smallest group is selected.

HDBSCAN labels points as noise if they do not fit well into the identified groups, but at the same time these points can include useful data points with significant log-likelihood values ($\ln P$).
To guarantee that the best log-likelihood values are correctly identified, we reassign such noise points to the relevant groups.
To achieve this, first, noise points (labeled -1) are identified.
If there are more than two unique groups (considering all excluding the noise points), the algorithm proceeds as follows: i) the group centers are determined by calculating the mean of the data points within each group; ii) the distances between each noise point and all group centers are determined; iii) if the distance to the nearest group center is less than the minimum inter-group distance\footnote{The inter-group distance is the minimum distance between the centers of the different groups.}, the noise points are reassigned to that group; iv) if there are fewer than two unique groups, all points, including noise points, are assigned to a single group.
This method ensures that significant data points, which may have been misclassified as noise due to their distance from dense regions, are accurately reassigned to appropriate groups.

\begin{figure}[htp]
\centerline{\includegraphics[width=0.5\textwidth]{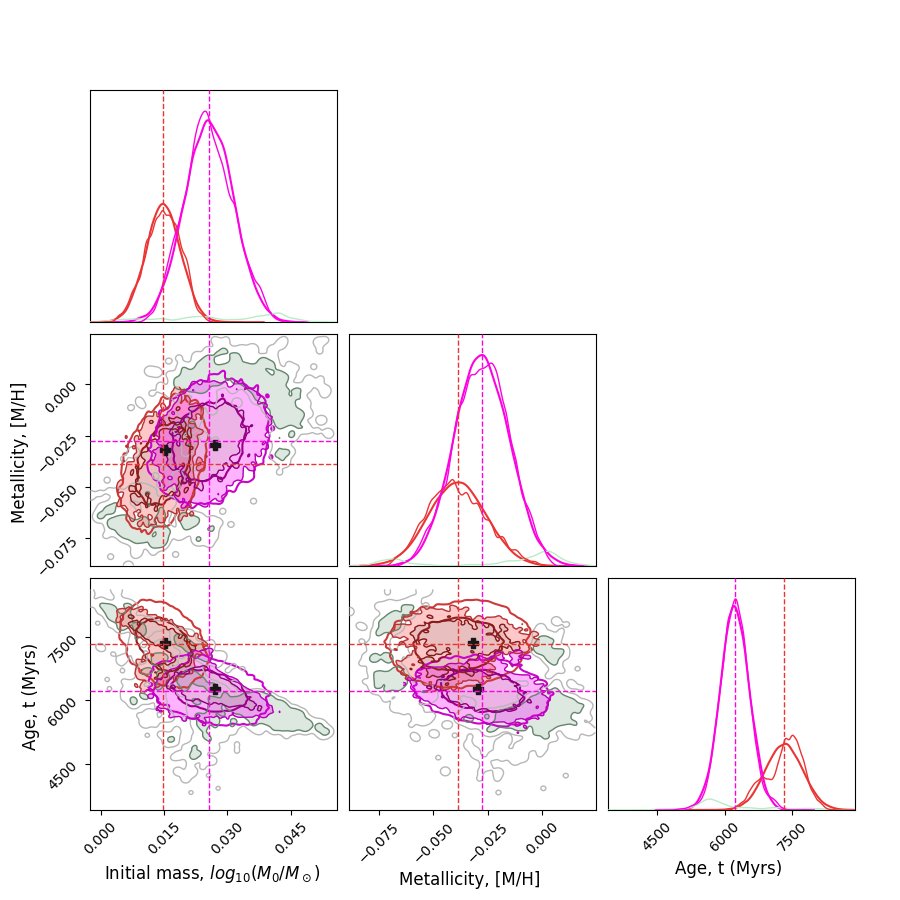}}
\caption{Triangle plot of the resulting PDF obtained by SPInS of a turnoff star of NGC 188, projected in the 3D space: \logM0, $\mathrm{[M/H]}$, $\mathrm{Age}$. The two groups of solutions identified by our algorithm are shown in red and magenta, whereas background noise is plotted in green.}
\label{fig:triangle}
\end{figure}

\begin{figure}[htp]
\centerline{\includegraphics[width=0.5\textwidth]{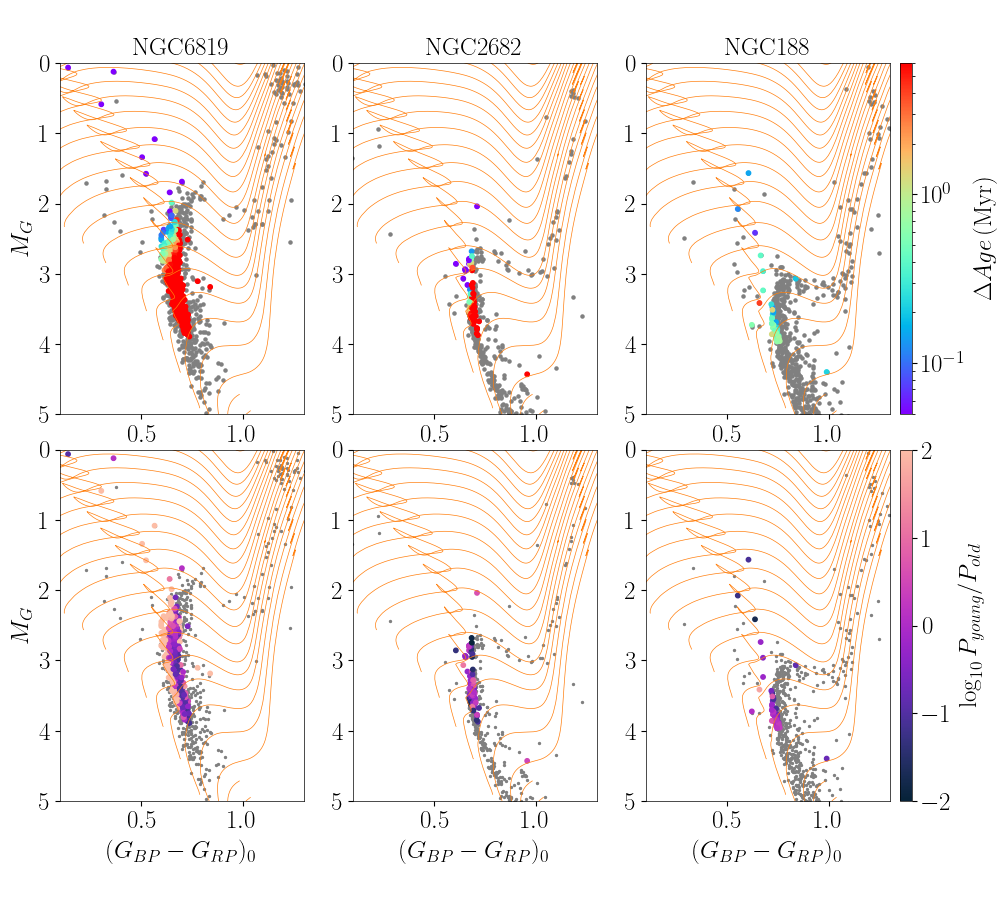}}
\caption{CMD of the three intermediate-age clusters for which we find the most multiple solutions.
Stars for which multiple peaks have been identified are plotted using coloured symbols according to the difference between the age of the most probable age peak and the mean literature age (top panels), and according to the probability ratio of the youngest to the oldest peak (bottom panels). Grey symbols depict stars with a single peak.
BaSTI stellar tracks of metallicity $-0.08$ are shown using orange lines to guide the eye.
}
\label{fig:multiplesols}
\end{figure}

\subsubsection*{Results obtained with clusters}
We detail here the results of applying the previous algorithm to the open and globular clusters results.
We have visually checked the performance of the algorithm for this smaller sample of stars (compared to the full sample of field stars Sect.~\ref{sec:MSTOSGB}), obtaining overall satisfactory results in the identification of multiple peaks.

From the total sample of 4,374 cluster stars (Sect.~\ref{sec:clustersel}) we find 813 stars with a multi-modal solution, such as the one depicted in Fig.~\ref{fig:triangle} for a turnoff star of NGC~188.
Of these, most of them have two solutions, while 42 stars have three solutions, and only 13 stars have more than three solutions.

We plot in Fig.~\ref{fig:multiplesols} the CMD of the stars with multiple solutions in the three intermediate-age clusters, for which we find the largest proportion of multi-modal distributions.
In the top panels the colour represents the difference between the age of the most probable PDF peak and the reference age taken as a mean of the literature ages listed in Table~\ref{tab:clusterdata}. 
In the bottom panels, we plot the ratio of the probabilities of the most probable peak vs the secondary peak.
BaSTI stellar tracks of around Solar metallicity (similar to the cluster's metallicities) are overplotted to guide the eye.

From Fig.~\ref{fig:multiplesols} we can see that the stars with multi-modal distributions are in general located along the turnoff loop, which is present around Solar metallicities for initial masses larger than $\sim1M_{\odot}$.
For smaller masses, the loop is not present any more and thus we do not find stars with multiple peaks in the low main sequence.
We also see clear gradients in the age difference of the main peak with respect to the cluster's real age (top panel, particularly for NGC~6819).
Finally, we see how the ratios of the probability of the youngest vs older peak gets more important in certain regions around the loop, particularly in the leftmost side of the MSTO for NGC~6819.

\subsection{Additional figures}\label{sec:appendix_fig}

\begin{figure}[htp]
\centerline{\includegraphics[width=0.5\textwidth]{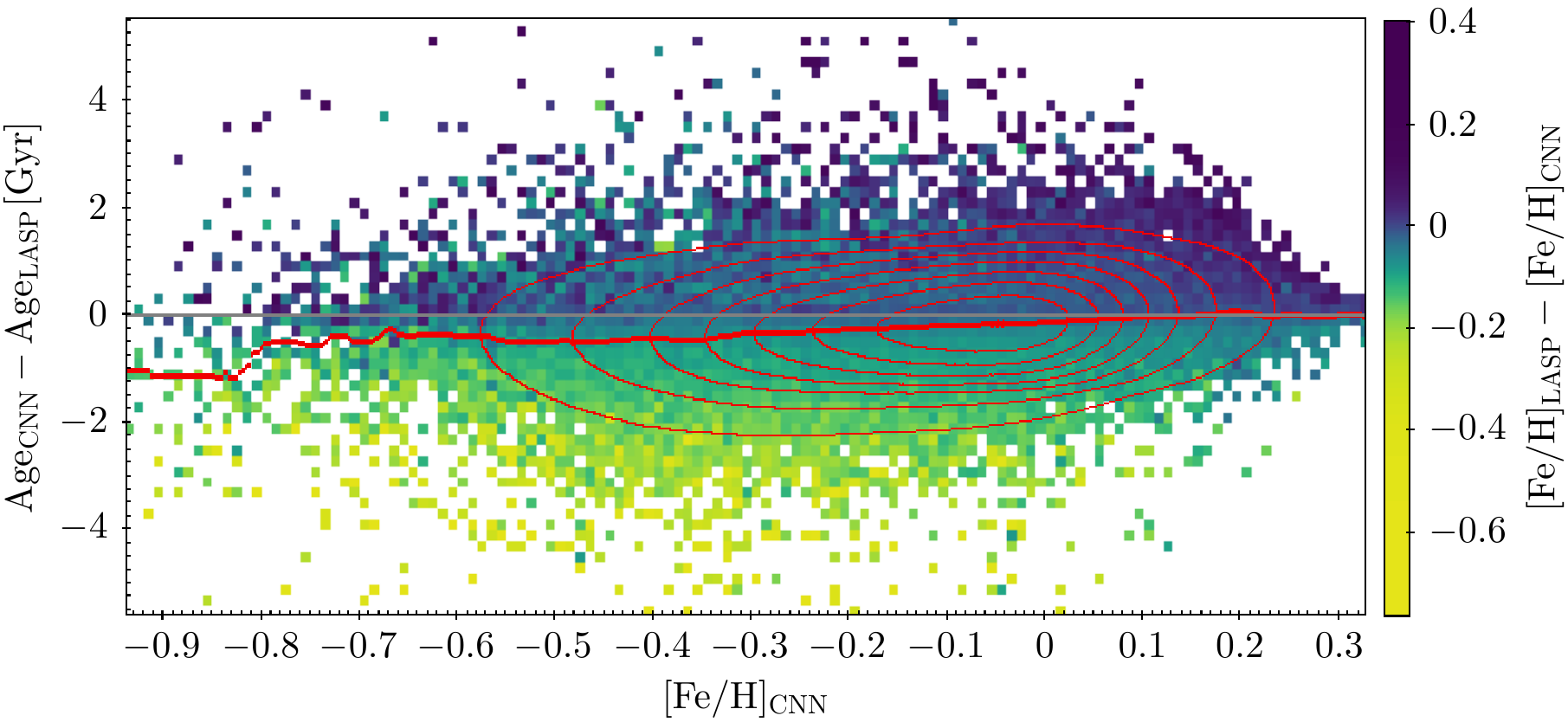}}
\caption{Age difference between LASP and CNN ages coloured by mean metallicity difference. Running median and contour lines are plotted in red.}
\label{fig:LASPCNN_diffage}
\end{figure}

We show in Fig.~\ref{fig:LASPCNN_diffage} the comparison between SPInS ages for the MRS obtained from LASP and CNN abundances, coloured by the mean difference in the metallicities.
A small trend towards metal-poor stars is present (mean value of -271 Myr), which correlates with the differences in the derived metallicities.

\begin{figure}[htp]
\centerline{\includegraphics[width=0.5\textwidth]{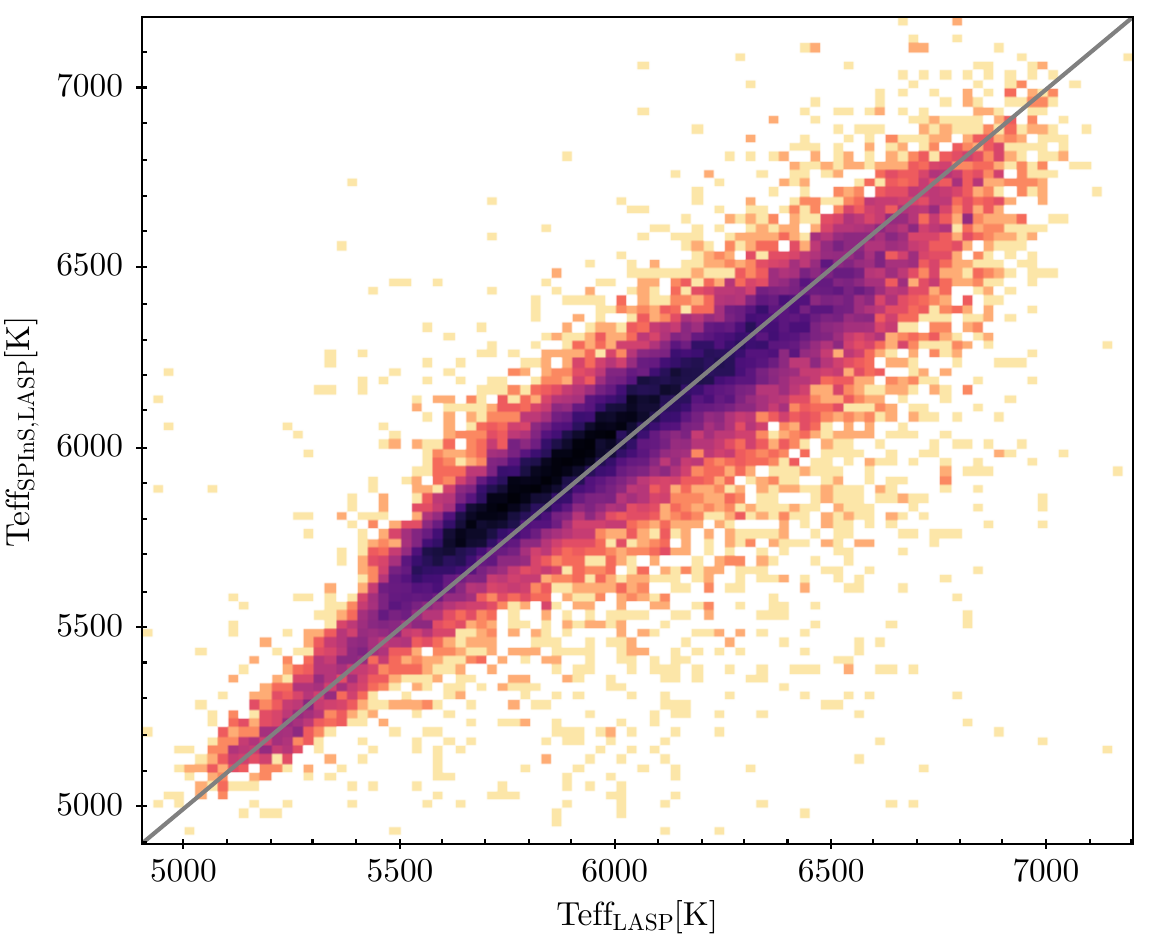}}
\centerline{\includegraphics[width=0.5\textwidth]{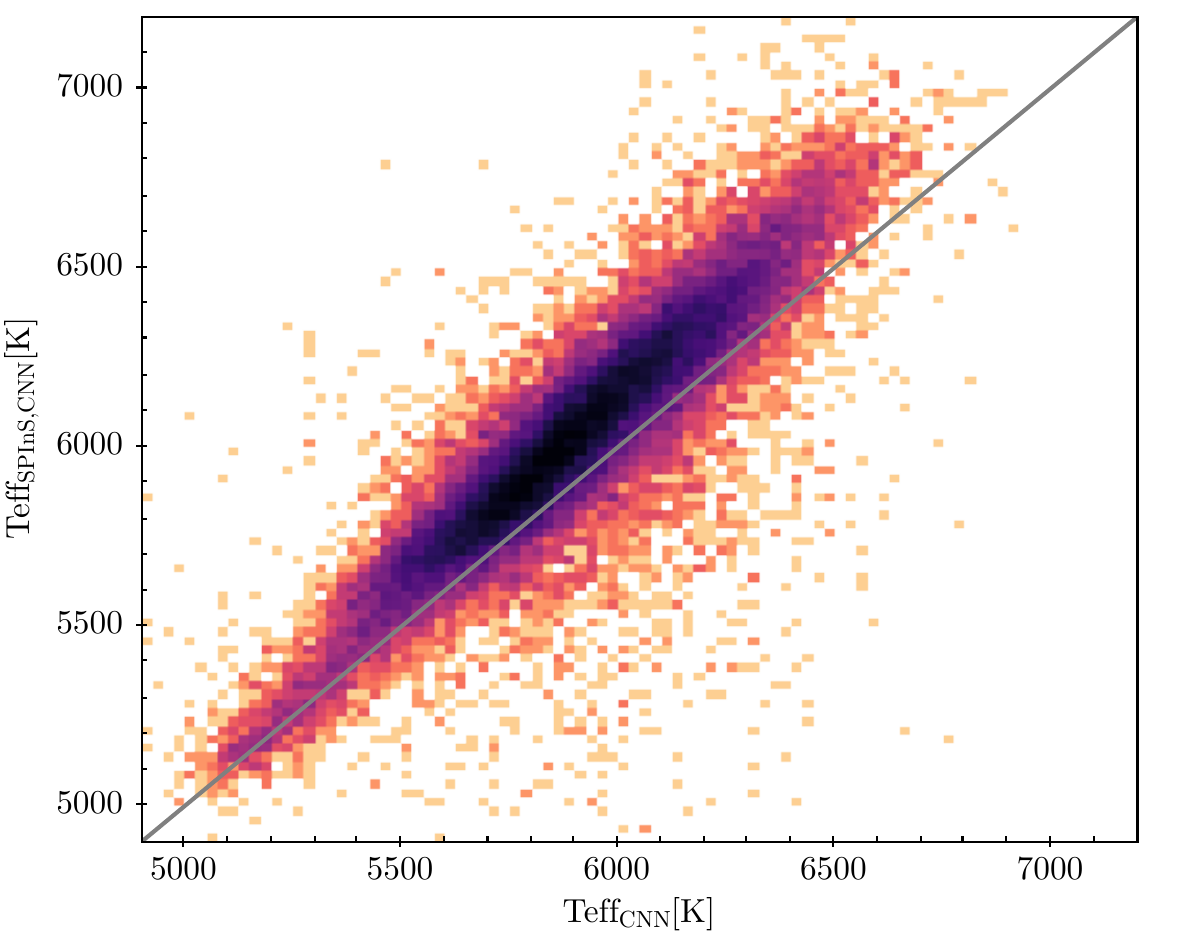}}
\caption{Spectroscopic determination of $\teff$ provided for the MRS by LASP (top) CNN (bottom), compared with the values obtained from the posterior distributions of SPInS.}
\label{fig:LASPCNN_Teff}
\end{figure}

Fig.~\ref{fig:LASPCNN_Teff} shows the $\teff$ comparison between the spectroscopic determination provided for the MRS by LASP/CNN, and the median value of the $\teff$ posterior distribution obtained by SPInS.
The obtained mean offsets and scatters are 23/166 K for LASP and 124/149 K for CNN, highlighting a discrepancy between the atmospheric parameters obtained by CNN with respect to the photometrically driven ones obtained by SPInS.

\begin{figure*}
    \centering
    \includegraphics[width=\textwidth]{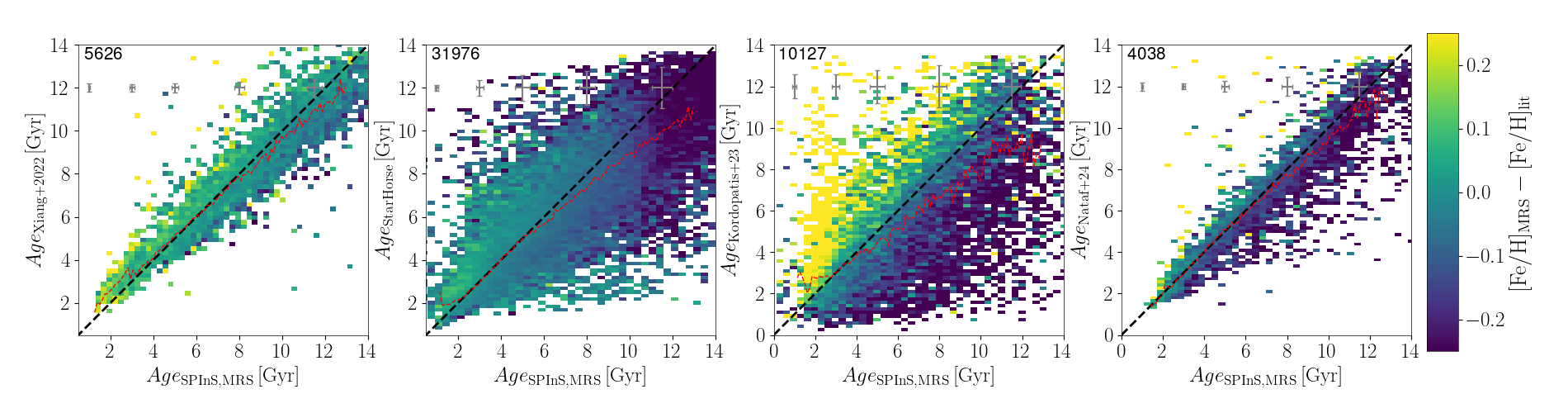}
    \includegraphics[width=\textwidth]{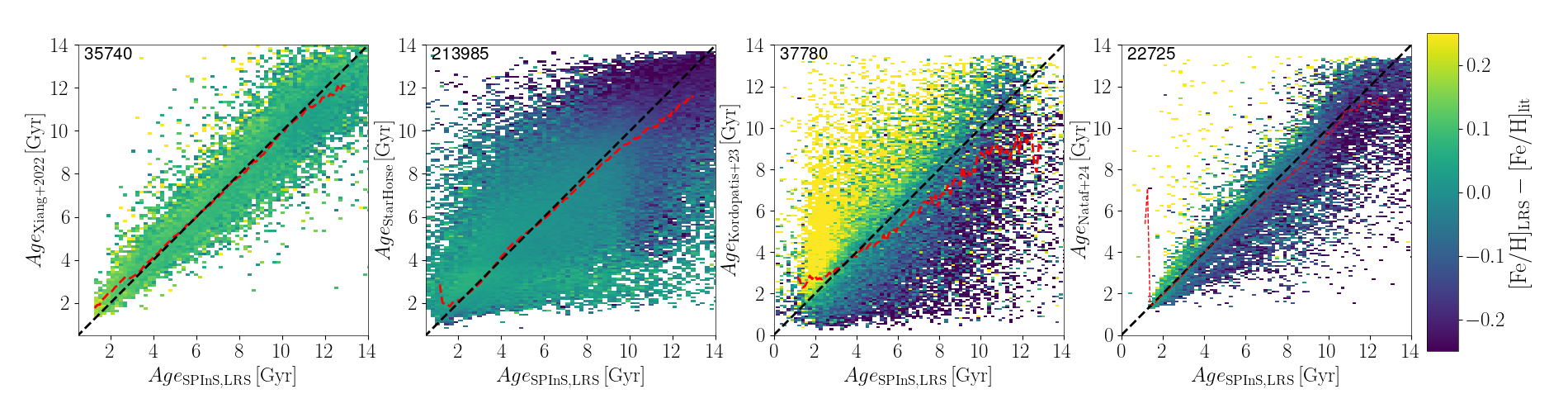}
    \caption{Comparison of the ages obtained for the MRS (top) and LRS (bottom) sample from SPInS with literature estimations estimates from \citet{Xiang+2022}, from StarHorse obtained in \citet{Queiroz+2023}, \citet{Kordopatis+2023} and \citet{Nataf+2024} (from left to right), coloured by mean difference in the metallicity from the two catalogues.
    The number of stars in common is indicated in each panel.
    A running median is shown in red dashed lines.}
    \label{fig:comp_met}
\end{figure*}

We show in Fig.~\ref{fig:comp_met} the comparison between literature and SPInS ages for the MRS and LRS coloured by the mean difference in the assumed metallicity between the two studies.
A running median is shown in red dashed lines.
See Fig.~\ref{fig:comparisonlit} for the same plots coloured by stellar density.
We see that the largest discrepancies between the derived ages are usually explained by the difference in the assumed metallicity, except for the catalog of \citet{Nataf+2024}, for which an overall offset of around 0.15 dex is obtained and no difference in age is present.

\end{document}